\documentclass[prd,twocolumn,showpacs,nofootinbib]{revtex4}
\usepackage{amsmath}
\usepackage{amssymb}
\usepackage{bm}
\usepackage{graphicx}
\begin{document}

\newcommand{\Rvir}{$R_{\mathrm{vir}}$}
\newcommand{\Rvire}{R_{\mathrm{vir}}}
\newcommand{\Mvir}{$M_{\mathrm{vir}}$}
\newcommand{\Mvire}{M_{\mathrm{vir}}}
\newcommand{\tdyn}{$t_{\mathrm{dyn}}$}
\newcommand{\tdyne}{t_{\mathrm{dyn}}}
\newcommand{\vk}{$v_{\mathrm{k}}$}
\newcommand{\vke}{v_{\mathrm{k}}}
\newcommand{\vvir}{$v_{\mathrm{vir}}$}
\newcommand{\vvire}{v_{\mathrm{vir}}}
\newcommand{\rhose}{\rho_{\mathrm{s}}}
\newcommand{\rse}{r_{\mathrm{s}}}
\newcommand{\Deltave}{\Delta_{\mathrm{v}}}
\newcommand{\rhoue}{\rho_{\mathrm{u}}}
\newcommand{\rhe}{r_{\mathrm{h}}}

\title{Mapping the allowed parameter space for decaying dark matter models}

\author{Annika H. G. Peter}
\email{apeter@astro.caltech.edu}
\affiliation{California Institute of Technology, Pasadena, CA 91125, USA}
\date{\today}

\begin{abstract}
I consider constraints on a phenomenological decaying dark matter model, in which two weakly interacting massive particle species have a small mass splitting, and in which the heavier particle decays to the lighter particle and a massless particle on cosmological time scales.  The decay parameter space is parameterized by \vk, the speed of the lighter particle in the center-of-mass frame of the heavier particle prior to decay, and the decay time $\tau$.  Since I consider the case in which dark matter halos have formed before there has been significant decay, I focus on the effects of decay on already formed halos.  I show that the $\vke-\tau$ parameter space may be constrained by observed properties of dark matter halos.  I highlight which set of observations is likely to yield the cleanest constraints on $\vke-\tau$ parameter space, and calculate the constraints in those cases in which the effect of decay on the observables can be calculated without $N$-body simulations of decaying dark matter.  I show that for $\vke \gtrsim 5\times 10^3\hbox{ km s}^{-1}$, the $z=0$ galaxy cluster mass function and halo mass-concentration relation constrain $\tau \gtrsim 40 \hbox{ Gyr}$, and that precise constraints on $\tau$ for smaller \vk~ will require $N$-body simulations.
\end{abstract}

\maketitle

\section{Introduction}\label{sec:intro}

Most of the mass energy of the Universe consists of at least two unknown but observationally substantiated materials (e.g., Ref. \cite{komatsu2009}).  While there is significant uncertainty as to what the ``dark energy'' that makes up $\sim 3/4$ of the Universe might be (for a review, see Refs. \cite{frieman2008,caldwell2009}), there are a number of attractive candidates for the $\sim 1/5$ of the Universe made of dark matter.  A popular class of dark matter candidate is the weakly interacting massive particle (WIMP), which includes the supersymmetric neutralino and the universal extra-dimensions Kaluza-Klein photon \cite{jungman1996,cheng2002}.  WIMPs are appealing because they appear naturally in extensions to the standard model (SM) of physics, are thermally produced in the early Universe in the quantity required by observations, and allow the large-scale structure of the Universe to evolve in a manner that appears consistent with observations (e.g., Refs. \cite{tegmark2004,lesgourgues2007,breid2009,komatsu2009,vikhlinin2009b}).

There are specific predictions on small scales, too.  The relatively large mass of WIMPs ($\gtrsim 100$ GeV) imply that thermal decoupling occurs when the WIMPs are nonrelativistic (``cold'') and with a high phase-space density.  The low speeds in combination with the typically weak scattering cross sections mean that only the smallest-scale perturbations are washed out by the time WIMPs kinetically decouple from light SM particles \cite{fillmore1984,hofmann2001,profumo2006,navarro2009}.  Hence, dark matter halos should be extremely dense at their centers and should exist down to $\sim$ Earth mass scales.  Moreover, dark matter density perturbations should evolve as a collisionless, pressureless fluid.

However, it is quite possible that dark matter consists of something other than standard WIMPs.  First, while observations are consistent with cold dark matter (CDM) on large scales (length scales of $\gtrsim 10$'s of Mpc), these observations do not \emph{require} a purely cold dark matter model.  Second, the interpretation of data on smaller scales is less clear or is nonexistent (e.g., Refs. \cite{dalcanton2001,simon2005}).  Third, WIMPs are by no means the only attractive class of particle candidates for dark matter. There are a number of theories in which the large-scale successes of cold dark matter are produced but deviate at the small scales where observations either do not exist or are more difficult to interpret \cite{spergel2000,feng2003,cembranos2005,profumo2005}.

In this work, I consider observational constraints on a phenomenological decaying dark matter model in which a parent particle $X$ experiences two-body decay of the form $X \rightarrow Y + \zeta$, where $Y$ is the stable daughter dark matter particle and $\zeta$ is a massless (or nearly massless) particle with extremely weak to nonexistent couplings with SM particles.  The $Y$ particle is only slightly less massive than the parent particle $X$, such that
\begin{eqnarray}\label{eq:eps}
	\epsilon = \frac{M_X - M_Y}{M_X} \ll 1.
\end{eqnarray}
For small $\epsilon$, the recoiling particle $Y$ receives a nonrelativistic speed 
\begin{eqnarray}\label{eq:vk}
	\vke /c = \epsilon
\end{eqnarray}
in the center-of-mass frame of the parent $X$ particle.  I consider large decay times $\tau \gtrsim 10^{12}$ s, such that decays occur after the first dark matter halos form in the early Universe, making this model similar to that considered in S{\'a}nchez-Salcedo (2003) \cite{sanchez2003}.  Such a model may arise in ``hidden sector'' theories, in which the SM does not interact with hidden parallel particle sectors (see Refs. \cite{feng2008,feng2008a,feng2009,ackerman2009} for examples of hidden sector theories).  Moreover, I assume that the parent $X$ particles are massive and thermally produced, such that the linear matter power spectrum in the early Universe should resemble that of CDM.

Since this model does not produce SM particles and the decay times are long, previously published constraints on decaying dark matter do not apply \cite{doroshkevich1984,ellis1985,kawasaki1995,feng2003a,chen2004,feng2004,kaplinghat2005,cembranos2007,cembranos2008,borzumati2008}.  Since the decay times I consider are long enough that dark matter halos should have formed, and formed with CDM-like properties, the main effect of decays is either to inject kinetic energy into or eject particles from existing dark matter halos.  The result of these effects is a change in the dark matter density profile in halos and halo mass loss.  Thus, I consider observational constraints on the model from the structure and abundances of dark matter halos.  In Sec. \ref{sec:vir}, I relate the two free parameters of the dark matter model, \vk~and $\tau$, to the typical speeds and time scales of particles in halos, and show how I expect halos to respond to decays as a function of the decay parameters.  In Sec. \ref{sec:obs}, I describe the types of observations that may constrain this decaying dark matter model, and how to translate observations into constraints on the decay parameters via the mapping in Sec. \ref{sec:vir}.  I calculate constraints in those instances in which they can be calculated semianalytically.  I summarize and discuss the results in Sec. \ref{sec:conclusion}.

\section{Relating Decay Parameters to Halo Parameters}\label{sec:vir}
The phenomenological decaying dark matter model is parameterized by just two numbers: the decay time $\tau$ and the kick speed (alternatively, the fractional mass difference between dark matter particles) \vk.  The effects of decay on dark matter halos as a function of these parameters depends on halo properties, namely the typical dynamical time (\tdyn) and the typical speed of dark matter particles in the halo.

The typical speed of dark matter particles in the halo is of order the virial speed,
\begin{eqnarray}
	\vvire = \sqrt{\frac{G\Mvire}{\Rvire}}, 
\end{eqnarray}
where the halo virial mass \Mvir~ is defined as the mass enclosed within the virial radius $\Rvire$ in which the average mass density is equal to the spherical collapse density $\Deltave \rhoue$,
\begin{eqnarray}
	\Mvire = \frac{4\pi}{3}\Deltave \rhoue \Rvire^3.
\end{eqnarray}
Here, $\rhoue$ is the mean mass density in the universe.  In a $\Lambda$CDM universe, $\Deltave(z) = (18\pi^2 - 39x - 82x^2)/\Omega_m(z)$, where $x = \Omega_m(z) - 1$ and $\Omega_m$ is fraction of the critical density  $\rho_\mathrm{c}$ in the form of matter ($\rhoue = \Omega_\mathrm{m} \rho_\mathrm{c}$) \cite{bryan1998}.

I take the typical dynamical time of a dark matter particle in a halo to be the crossing time at the half-mass radius of the halo.  I do not use the crossing time at the virial radius because 
\begin{eqnarray}
	t_{\mathrm{vir}} &\sim& \vvire/\Rvire \\
		&\sim& \sqrt{\Mvire/\Rvire}/\Rvire \\
		&\sim& \Rvire/\Rvire = \hbox{ const},
\end{eqnarray}
which is the same for all halos.  I must select a halo model in order to calculate the dynamical time as defined as the crossing time at the half-mass radius.  Dissipationless cosmological simulations of cold dark matter alone (without baryons) show that dark matter halo profiles are well described by the Navarro-Frenk-White (NFW) profile
\begin{eqnarray}
	\rho(r) = \frac{\rhose}{\frac{\displaystyle r}{\displaystyle \rse}\left( 1 + \frac{\displaystyle r}{\displaystyle \rse} \right)^2} \label{eq:nfw}
\end{eqnarray}
on observable scales (radii from galactic centers $r > 0.01\Rvire$) \cite{navarro1997,navarro2004,navarro2009}.  The scale radius $\rse$ can be related to the $\Rvire$ by
\begin{eqnarray}
	\rse = \Rvire / c, \label{eq:rse}
\end{eqnarray}
where $c$ is the halo concentration.  The concentration is expected to be a function of the formation time of the halo, and is thus in general a function of the mass of the halo \cite{bullock2001,lu2006,neto2007}.  The scale density $\rhose$ can be related to virial quantities,
\begin{eqnarray}
	\rhose = \frac{\Deltave \rhoue}{3}\frac{c^3}{\ln(1 + c) - c/(1+c)}.
\end{eqnarray}

If $c\gg 1$, which is expected for virialized halos, then the half-mass radius 
\begin{eqnarray}
	\frac{\rhe}{\rse} \approx 1.65 \sqrt{c}.
\end{eqnarray}
The typical dark matter particle speed at such a radius is typically
\begin{eqnarray}
	v_{\mathrm{h}} &\sim& \sqrt{\frac{G\Mvire}{2\rhe}} \\ 
		       &\sim& \left( G\Deltave \rhoue \right)^{1/2} c^{1/4} \Rvire,
\end{eqnarray}
which is similar to the virial speed.  Thus, the typical crossing speed at the half-mass radius, and hence, the typical dynamical time, is
\begin{eqnarray}
	\tdyne &\sim& \rhe/v_{\mathrm{h}} \\
		&\approx& (G \Deltave \rhoue )^{-1/2} c^{-3/4}.
\end{eqnarray}
In CDM, a typical galaxy ($\Mvire \sim 10^{12} M_\odot$) has $c \approx 15$, yielding $\vvire = 130 \hbox{ km s}^{-1}$ and $\tdyne = 500$ Myr; a large galaxy cluster has $\Mvire \sim 10^{15} M_\odot$ and $c\sim 5$, corresponding to $\vvire = 1300 \hbox{ km s}^{-1}$ and $\tdyne \sim $ Gyr; and the ultra-faint dwarf galaxies in the Local Group likely have virial masses $\sim 10^9 M_\odot$ ($\vvire = 13\hbox{ km s}^{-1}$) \cite{strigari_nat2008,bullock2001,maccio2007,neto2007,maccio2008}.  The concentration of those galaxies is unknown, but taking $c = 30$ leads to $\tdyne = 200$ Myr.

Another interesting time scale is the crossing time at the scale radius, $t_{\mathrm{s}}$, since this corresponds to the region of the halo at which the profile transitions from $\rho \propto r^{-1}$ to $\rho \propto r^{-3}$.  The ``scale time'' $t_{\mathrm{s}}$ is $\sim 50$ Myr for an ultra-faint dwarf galaxy halo, $\sim 100$ Myr for a typical galaxy halo, and $\sim 400$ Myr for the cluster-sized halo.  This highlights the point that although I choose one time scale to parameterize the dynamical time in a halo, there is actually a diversity of dynamical times within a halo.

Now that I have parameterized the time scales and speeds of dark matter particles in halos, I will classify the effects of dark matter decays are on halos as a function of the decay parameters with respect to the halo parameters.  At the present, I will ignore cosmological accretion onto halos, and consider the halos to be isolated and in equilibrium.

\emph{Case 1} ($\tau > \tdyne,~\vke > \vvire$)---This is a regime in which the decay time is long relative to the dynamical time of the halo, and the kick speed is high.  The limit of relativistic kicks was studied in Refs. \cite{flores1986} and \cite{cen2001} in other contexts, but even for nonrelativistic kicks greater than \vvir, the $Y$ particles are ejected from the system.  The decays do not directly inject kinetic energy into the bound halo, as they will in the following cases, because the decay products $Y$ will be unbound to the halo.

In cases in which the time scale for change in the gravitational potential is significantly longer than the dynamical time, particles on regular orbits should conserve adiabatic invariants.  This is useful because, if the halo is quasi-static, the distribution function (DF) of dark matter particles is a function of adiabatic invariants.  The mass density in the halo thus be calculated using this DF if the gravitational potential of the halo is known (cf. Ref. \cite{binney2008}).  In the simplified case of a spherically symmetric potential with all particles initially on circular orbits, the gravitational potential, and hence, the dark matter density of the halo may easily be found as a function of the fraction of particles that have been ejected from the halo.  The approximation of dark matter particles on circular orbits is not as unrealistic as one might expect; cosmological simulations of dark matter halos show that there is not a significant phase-space density of highly radial orbits, except in the outskirts of the halo \cite{diemand2009}.  This approximation was used in Refs. \cite{flores1986} and \cite{cen2001}, and I show analytically how the density profile of dark matter halos changes as a function of the fraction of the $X$ particles that have decayed, $f$.  

If the angular momentum of particles is conserved, then the initial and final mass distributions are related as
\begin{eqnarray}
	M_\mathrm{i}(r_\mathrm{i})r_\mathrm{i} = M_\mathrm{f}(r_\mathrm{f}) r_\mathrm{f} \label{eq:j_conservation},
\end{eqnarray}
since the specific angular momentum of particles on circular orbits is $J = (GM(r)r)^{1/2}$, where $M(r)$ is the mass enclosed within radius $r$.  Here, i denotes the initial halo properties, and f denotes halo properties after a fraction $f$ of $X$ particles have decayed.  If the $X$ particles are initially on circular orbits, then their orbits do not cross, allowing for the following relation:
\begin{eqnarray}
	M_\mathrm{f}(r_\mathrm{f}) = (1-f)M_\mathrm{i}(r_\mathrm{i}).
\end{eqnarray}
Inserting this equation into Eq. (\ref{eq:j_conservation}), I find that the initial and final particle radii are related as
\begin{eqnarray}
	r_\mathrm{i} = (1-f)r_\mathrm{f},
\end{eqnarray}
such that
\begin{eqnarray}
  M_\mathrm{f}(r_\mathrm{f}) = (1-f)M_\mathrm{i}((1-f)r_\mathrm{f}).
\end{eqnarray}
This changes the mass density of the $X$ particles in the following way.  The density is
\begin{eqnarray}
  \rho_\mathrm{f}(r_\mathrm{f}) &=& \frac{1}{4\pi r_\mathrm{f}^2} \frac{\displaystyle \text{d}M_\mathrm{f}}{\displaystyle \text{d}r_\mathrm{f}} \\
  &=& \frac{1}{4\pi r_\mathrm{f}^2} (1-f) \frac{\text{d}M_\mathrm{i}}{\text{d}r_\mathrm{f}} \\
  &=& \frac{(1-f)^2}{4\pi r_\mathrm{f}^2} \frac{\text{d}M_\mathrm{i}}{\text{d}r_\mathrm{i}}.
\end{eqnarray}
In the case of an NFW halo [Eq. (\ref{eq:nfw})], this implies
\begin{eqnarray}
  \rho_\mathrm{f}(r_\mathrm{f}) = \frac{(1-f)^4 \rhose}{\left( \frac{\displaystyle (1-f)r_\mathrm{f}}{\displaystyle \rse} \right) \left[1 + \frac{\displaystyle (1-f)r_\mathrm{f}}{\displaystyle \rse} \right]^2 }.
\end{eqnarray}
Thus, the halo retains its NFW form, but with scale radius 
\begin{eqnarray}
	r_\mathrm{s,f} = \left( 1 - f \right)^{-1}r_\mathrm{s,i}, \label{eq:rse_new}
\end{eqnarray}
and with a decreased scale density
\begin{eqnarray}
	\rho_\mathrm{s,f} = \left( 1 - f \right)^4 \rho_\mathrm{s,i}. \label{eq:rhose_new}
\end{eqnarray}

Thus, if $\tau \gg \tdyne$ and $\vke \gg \vvire$, the \emph{shape} of the dark matter halo will be unchanged, but the scale radius will increase, and the mass density and total virial mass will decrease.  Note that these results are independent of \vk~ beyond the fact \vk~ must be large enough to unbind any $Y$ particle from the halo.

\emph{Case 2} ($\tau > \tdyne,~\vke < \vvire$)---In this regime, the halo is slowly evolving as a function of time, but the kick speeds are small.  Unlike case 1 (above), it is difficult to calculate analytically the general behavior of the halo.  Although the decay time scale is long enough that the gravitational potential of the halo should evolve fairly slowly, the $Y$ daughter particles largely stay within the halo, making it difficult to estimate changes to the adiabatic invariants or to the gravitational potential.  

However, there are a few general predictions one may make.  First, since self-gravitating systems have negative heat capacity, any slow injection of kinetic energy into the halo causes the halo to expand and for the typical particle energy to become less negative (see, e.g., Ref. \cite{binney2008}).  Second, there will be some mass loss as some initially loosely bound particles will decay to particles that are no longer bound to the halo, and more may be lost as the gravitational potential responds to the kinetic energy injection resulting from decays.  The consequences of these effects are to drive down the central halo density, the total virial mass, and the typical particle speed.  The effects will be larger for high \vk~and shorter $\tau$, and will be most pronounced at the halo centers (where the typical particle speed is smaller).  For small \vk~or large $\tau$, there should hardly be a change to halo properties.

In general, the behavior of decays in this regime must be examined either by solutions to the Boltzmann and Poisson equations or by $N$-body simulations.  

\emph{Case 3} ($\tau < \tdyne,~\vke > \vvire$)---This regime, in which the decay time is less than the dynamical time, can be thought of as being similar to the case of instantaneous decay.  In this particular case, the kick speed is also quite high.  In general, most, if not all, of the mass in the halo will be ejected.  If any mass remains in the halo, the system will settle to a new equilibrium within several dynamical times, but it is not clear what the structure of that halo will be.  In the absence of accretion, the structure of the halo is fixed for the rest of time.  

Again, quantitative predictions for this regime require $N$-body simulations.

\emph{Case 4} ($\tau < \tdyne,~\vke < \vvire$)---In this case, the velocity perturbations are small and occur in one short epoch.  This is analogous to the case of high-speed galaxy encounters.  And like noncatastrophic high-speed galaxy encounters, we can calculate one post-encounter (or post-decay) property of the halo: the total kinetic energy in the virial radius.  If a halo is initially in virial equilibrium, the total energy $E$ is related to the kinetic energy $K$ by (cf. Ref. \cite{binney2008}):
\begin{eqnarray}
	E = - K.
\end{eqnarray}
If no particles are ejected as a result of the decays, the total kinetic energy injected in the halo is $\Delta K = M_{\mathrm{vir}} \vke^2/2$.  Thus, the final energy will be
\begin{eqnarray}
	E_\mathrm{f} = - K + \Delta K,
\end{eqnarray}
and after the halo has settled into virial equilibrium again after a few dynamical times, the kinetic energy will be
\begin{eqnarray}
	K_\mathrm{f} &=& - E_\mathrm{f}\\
		     &=& K - \Delta K.
\end{eqnarray}
Once again, the negative heat capacity of self-gravitating systems means that the net effect of injection of kinetic energy into a halo is a decrease in the total kinetic energy of the system.

If \vk~is a significant fraction of \vvir, this approximation for the final kinetic energy becomes worse, as decays result in mass loss from the system, both directly as individual $Y$ particles are created on unbound orbits, and indirectly due to the rapidly changing gravitational potential.

Even if the final energy of the halo is known, the form of the gravitational potential (and hence, the density distribution) is not without the help of $N$-body simulations.  However, just as in case 2, we predict that the halo will become less dense, and that the mass loss and density changes will be more extreme for larger values of \vk.

There are several points to make about these regimes.  First, given that galaxies and clusters span a large range of halo mass, a fixed $\tau$ and $\vke$~will place a low mass halo in a different regime than a high mass halo.  For example, if $\tau \sim 1$ Gyr and $v\sim 100$ km s$^{-1}$,  ultra-faint dwarf galaxies will be in case 1.  They will have disintegrated by the present since particles will be ejected from the halo on time scales far less than the age of the universe, $t_\mathrm{H} \approx 14$ Gyr.  However, the halo of an typical ($\sim 10^{12}M_\odot$) galaxy will be in the regime of case 2, and the halo of massive cluster will be case 4.  Secondly, there will be a differentiation in cases throughout a single halo since dynamical time scales are much shorter deep within the halo.  A halo may be in case 2 in the interior, but in case 4 near the edge of the halo.  Last, while one can make general predictions for the behavior of halos for decays in each regime, quantitative predictions largely require $N$-body simulations, which we perform and analyze in another paper \cite{peter2010}.

\begin{figure}
	\includegraphics[height=0.45\textwidth,angle=270]{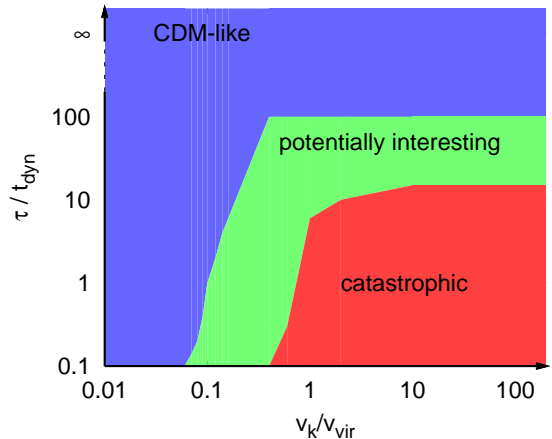}
	\caption{\label{fig:vir}Summary of the changes to dark matter halos due to late-time decays, as a function of \vk/\vvir~and $\tau$/\tdyn.  In the region marked ``CDM-like'', decays have little effect on halos, and so halos should resemble those produced by CDM.  The area marked ``catastrophic'' is the region of parameter space in which halos are almost completely destroyed as a result of decays.  The middle region labeled ``potentially interesting'' indicates the part of parameter space that is not ruled out as catastrophic, nor is it likely to produce halos that exactly resemble those predicted by CDM.}
\end{figure}

The results are summarized in Fig. \ref{fig:vir}.  The region marked ``CDM-like'' denotes the $\tau \gg \tdyne$ and $\vke \ll \vvire$ parts of parameter space in which we expect either almost no decays or almost no kinetic energy injection in the halo.  For this region of parameter space, the halo properties will be nearly indistinguishable from the case of no decays.  The region marked ``catastrophic'' has extended beyond the parts of parameter space of case 3 to lower \vk~ and higher $\tau$, since $\vke \sim \vvire$ can still induce significant mass loss, and decays on time scales of a few dynamical times can affect the halo as it settles into its new equilibrium.  The region marked ``potentially interesting'' includes bits of cases 1, 2, and 4, and represents the part of parameter space whose effects on halo structure need to be explored with $N$-body simulations and constrained with observations.  In the next section, I show how to relate observational constraints to Fig. \ref{fig:vir}, and define regions of $\vke-\tau$ parameter space allowed by observations and which need to be better understood with simulations.

\section{Observational Constraints}\label{sec:obs}
There are a number of probes of the distribution of matter in the Universe.  In this section, I describe a set of probes from which constraints on decaying dark matter are easiest to infer, and calculate constraints for case 1-type decays.  We consider observational constraints on case 2, 3, and 4 decays in other work \cite{peter2010}.  The probes I consider are the cluster mass function (Sec. \ref{sec:obs:cluster}), galaxy clustering (Sec. \ref{sec:obs:pofk}), the existence of small dark matter halos (Sec. \ref{sec:obs:small}) and the mass-concentration relation (Sec. \ref{sec:obs:mc}).

\subsection{The Cluster Mass Function}\label{sec:obs:cluster}
The cluster mass functions are relatively clean for decaying dark matter studies for both observational and theoretical reasons.  Individual cluster halo masses are substantially easier to determine than the halos of individual galaxies, and the assignment of a cluster of galaxies to a single dark matter halo is unambiguous.  Moreover, since most of the energy in decays goes to nonrelativistic massive particles, the background evolution of the Universe is unchanged.  Hence, the growth function, which depends only on redshift and the background cosmology in a $\Lambda$CDM universe, is unchanged relative to $\Lambda$CDM predictions (see Refs. \cite{oguri2003,gong2008} for the relativistic case, in which the background evolution does change) \cite{dodelson2003}.  Thus, main change to the cluster mass function relative to that predicted from $\Lambda$CDM models occur as a result mass loss from the halos due to decay.  This makes the mass functions simple to interpret with respect to $\Lambda$CDM predictions.

There are two different ways to use the cluster mass function to constrain the decaying dark matter parameter space.  I will show how the $z=0$ mass function (see Refs. \cite{bahcall2003a,rines2007,rines2008b,rozo2010,vikhlinin2009a,vikhlinin2009b} for recent observations and mass-function calculations) constrains the decay time $\tau$ in the case 1 virial regime.  This method is easy to generalize to other parts of decay parameter space.  Second, I show how upcoming cluster surveys (e.g., Sunyaev-Zel'dovich observations in the microwave bands \cite{carlstrom2002}) that are sensitive to the redshift evolution of the cluster mass function can be used to constrain decays.  

\subsubsection{The local cluster mass function}
Cluster mass functions derived from X-ray and optical surveys of the local Universe have been used to constrain $\Omega_m$ and $\sigma_8$, the rms amplitude of fluctuations in the linear density field at $z=0$ if smoothed on 8 $h^{-1}$ Mpc scales, where $h$ is the Hubble parameter \cite{rines2008b,rozo2010,vikhlinin2009b}.  The constraints on $\Omega_m$ and $\sigma_8$ from the local cluster mass function are consistent with those found with other cosmological probes, including the cosmic microwave background (CMB) \cite{dunkley2009}, the Lyman-alpha forest \cite{lesgourgues2007}, weak-lensing power spectra, and galaxy clustering \cite{heymans2005,seljak2005,tegmark2006,mandelbaum2007,massey2007,padmanabhan2007,percival2007,breid2009}.

The consistency of the constraints on $\Omega_m$ and $\sigma_8$ from different epochs constrains decaying dark matter for the following reason.  The standard lore of halo formation is that halos form when the amplitude of the density perturbation (alternatively, the amplitude of the linear matter power spectrum) smoothed on a distance scale $R$ (or alternatively, a mass scale $M = 4\pi \rho_u R^3 / 3$), exceeds the overdensity required for halo collapse \cite{press1974,sheth2002}.  The halo number density $dn/dM$ is then related to the volume of space in which the overdensity exceeds the collapse density.  If the smoothed amplitude of fluctuations is small, then there are few virialized halos of that given mass; if the amplitude is large, then there are many halos of a given mass.  The amplitude of the linear matter power spectrum is controlled by the initial amplitude of fluctuations from inflation as well as $\Omega_m$, but is usually parameterized in terms of $\sigma_8$ and $\Omega_m$.  

The number density of clusters, the rarest of virialized objects, is quite sensitive to $\sigma_8$ and $\Omega_m$, with more clusters expected in a high-$\sigma_8$, high-$\Omega_m$ cosmology.  A $\sim 20\%$ change in $\sigma_8$ can yield order-unity changes in the cluster halo number density (e.g., Ref. \cite{tinker2008}).  Decays cause mass loss in halos, thus reducing the number density of halos above fixed mass.  If one were to infer $\Lambda$CDM parameters from the local cluster mass function if dark matter were decaying, one would infer artificially small values of $\sigma_8$ and $\Omega_m$ relative to what one would infer from probes of earlier epochs (e.g., the CMB and Lyman-alpha forest).  The fact that $\sigma_8$ and $\Omega_m$ inferred from the local cluster mass function are similar to those inferred from other probes limits decay parameter space, although the error bars for the cluster mass functions (and hence, those for the inferred values of $\sigma_8$ and $\Omega_m$) are quite large \cite{rines2008b,rozo2010,vikhlinin2009b}.

One can estimate conservative limits on decay parameter space using the following method.  Since decays lower the inferred $\sigma_8$ and $\Omega_m$ of clusters relative to those inferred from the CMB, I find which combination of \vk~ and $\tau$ would cause the 2-$\sigma$ upper limits on $\Omega_m$ and $\sigma_8$ from the CMB to yield a $z=0$ cluster mass function that is barely consistent to the 2-$\sigma$ lower limits on those parameters.  In the instance of case 1 of Sec. \ref{sec:vir}, this can be done analytically.  Here, I show the limits one can set on $\tau$ assuming that \vk~ is several times the typical cluster virial speed ($\vvire \sim 10^3\hbox{ km s}^{-1}$).

\begin{figure}
	\includegraphics[width=0.45\textwidth]{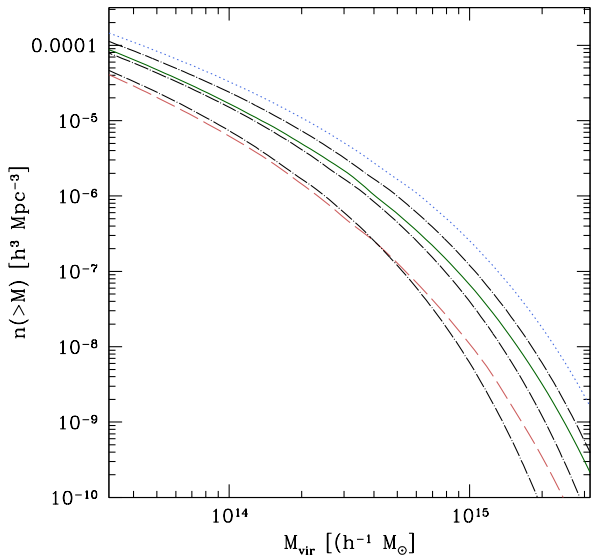}
	\caption{\label{fig:ngreater}Halo number density $n(>M)$ as a function of minimum halo mass $M$ at $z=0$.  The dotted (blue) line represents the number density of halos in a cosmology with $\Omega_m = 0.318$ and $\sigma_8 = 0.868$ (with both quantities being $2$-$\sigma$ above the WMAP-5 mean values), the solid (green) line represents the halo number density for a WMAP-5 cosmology ($\Omega_m = 0.258$, $\sigma_8 = 0.796$), and the dashed (red) line shows the number density of halos in a cosmology with $\Omega_m$ (=0.198) and $\sigma_8$ (=0.724) $2$-$\sigma$ below the WMAP-5 mean values.  These number densities were generated using the Tinker et al. \cite{tinker2008} mass-function fitting formulas.  The dot-dashed lines show halo number densities (top to bottom) once a fraction $f = 0.1, 0.2,$ and 0.3 of the $X$ particles have decayed to fast $Y$ particles, assuming a high-$\Omega_m$, high-$\sigma_8$ cosmology.}
\end{figure}

I take the $\Omega_m$ and $\sigma_8$ and the associated error bars from the \emph{Wilkinson Microwave Anisotropy Probe} (WMAP) five-year data set as the baseline, since this is the earliest epoch from which $\Omega_m$ and the amplitude of the matter power spectrum can be inferred \cite{dunkley2009}.  The central values and 1-$\sigma$ error bars from the WMAP-5 6-parameter $\Lambda$CDM fits are $\Omega_m = 0.258 \pm 0.030$ and $\sigma_8 = 0.796 \pm 0.036$.  I use the Tinker et al. (2008) \cite{tinker2008} mass function to estimate the comoving number density $n(>M)$ of halos above a mass threshold $M$ at $z=0$ for $\Omega_m = 0.318$ and $\sigma_8 = 0.868$ (2-$\sigma$ above the WMAP-5 mean values) and $\Omega_m = 0.198$ and $\sigma_8 = 0.724$ (2-$\sigma$ below), keeping the Hubble parameter $h$, the primordial slope of the matter power spectrum $n_s$, and the baryon fraction $\Omega_b h^2$ fixed to the WMAP-5 central values, assuming a flat $\Lambda$CDM cosmology.  The cluster mass function is far less sensitive to those parameters than to $\Omega_m$ and $\sigma_8$.  The mass functions are shown in Fig. \ref{fig:ngreater}.  

Then, assuming there is a one-to-one mapping between a $\Lambda$CDM halo virial mass $M_\mathrm{i}$ (calculated using the spherical top-hat overdensity described in Sec. \ref{sec:vir}) for the high-$\Omega_m$, high-$\sigma_8$ cosmology and the virial mass $M_\mathrm{f}$ after a fraction $f$ of the $X$ particles in the halo have decayed, I map the high-$\Omega_m$, high-$\sigma_8$ $\Lambda$CDM comoving number density $n_\mathrm{i}$ to the comoving number density of halos after a fraction $f$ of $X$ particles have decayed
\begin{eqnarray}
	n_\mathrm{f} (> M_\mathrm{f} ) = \int_{M_{\mathrm{f}}}^\infty dM_\mathrm{f}^\prime \frac{dn_\mathrm{i}(M_{\mathrm{i}}(M^\prime_\mathrm{f}))}{dM_\mathrm{i}} \frac{dM_{\mathrm{i}}}{dM_\mathrm{f}^\prime}.
\end{eqnarray}
The resulting comoving number densities are shown as the dashed lines in Fig. \ref{fig:ngreater} for $f=0.1,0.2$ and $0.3$.

The comoving number density of cluster halos with $M_\mathrm{vir} > 10^{14} h^{-1}M_\odot$ (roughly the mass threshold for cluster studies) if $f=0.3$, corresponding to $\tau \approx 40$ Gyr, is nearly identical to that of the low-$\Omega_m$, low-$\sigma_8$ cosmology, although the mass function has a steeper slope at larger masses $\sim 5\times 10^{14}h^{-1}M_\odot$.  For slightly smaller values of $f$ ($\approx 0.25$), the decay mass function lies above the low-$\Omega_m$, low-$\sigma_8$ mass function for $M_\mathrm{vir}\lesssim 10^{15} h^{-1}M_\odot$.  Thus, the upper limit on $f$ is near $0.3$, such that $\tau \gtrsim 40$ Gyr for $\vke >$ a few times the cluster virial speed of the most massive observed clusters.  Since clusters are observed to have $\Mvire \lesssim 2\times 10^{15} M_\odot$, this calculation applies to $\vke \gtrsim 5000\hbox{ km s}^{-1}$.

If the decay parameters relative to the cluster dynamical parameters are in any case other than case 1 of Sec. \ref{sec:vir}, one will need to perform $N$-body simulations of decay in halos in order to map decay parameters to the halo mass loss.

The $f=0.3$, $\tau \approx 40$ Gyr lower limit for case 1 is likely to be too conservative, since I have assumed a one-to-one mapping between a $\Lambda$CDM halo of mass $M_\mathrm{i}$ and a halo in the decay case with mass $M_\mathrm{f}$.  However, to do this sort of analysis self-consistently, one should perform cosmological $N$-body simulations to take into account the streaming of dark matter particles.

\subsubsection{Redshift evolution of the mass function}
The evolution of the cluster mass function is a promising way in which to explore evolution in the dark-energy equation of state $w$, since the evolution in the comoving number density of cluster mass halos is quite sensitive to the growth function and to the Hubble constant as a function of redshift, $H(z)$.  While current constraints on $w$ from cluster mass functions are weak \cite{vikhlinin2009b}, ongoing and future Sunyaev-Zel'dovich surveys and follow-up observations should provide much better constraints (e.g., Ref. \cite{carlstrom2002,cunha2009}).

\begin{figure}
	\includegraphics[width=0.45\textwidth]{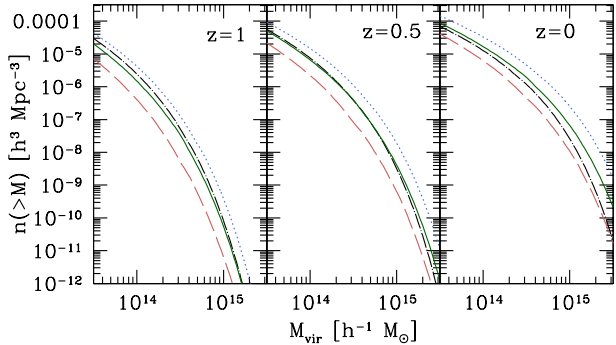}
	\caption{\label{fig:nevolve}Evolution of the number density of massive halos as a function of time.  The lines have the same meaning as in Fig. \ref{fig:ngreater}, and the dot-dashed line represents a high-$\Omega_m$, high-$\sigma_8$ cosmological model with $\vke > \vvire$ of clusters and $\tau = 50$ Gyr.}
\end{figure}

The same reasons why the evolution of the cluster mass function is a useful probe of dark energy also make it a useful probe of decays, which was originally pointed out in Refs. \cite{cen2001} and \cite{oguri2003}.  In the right-hand panel of Fig. \ref{fig:nevolve}, I show the $z=0$ comoving number densities of halos for the high-$\Omega_m$, high-$\sigma_8$; mean WMAP-5; and low-$\Omega_m$, low-$\sigma_8$ cosmologies, as well as the comoving number density of halos if $\tau = 50$ Gyr in case 1 of Sec. \ref{sec:vir}, $\Omega_m = 0.318$, and $\sigma_8 = 0.868$.  The middle and left panels show the comoving number densities at $z=0.5$ and $z=1$, respectively.  One can see that the $z=0$ mass function for $\tau = 50$ Gyr lies between the low-$\Omega_m$, low-$\sigma_8$ and mean WMAP-5 mass functions, but it lies between the mean WMAP-5 and high-$\Omega_m$, high-$\sigma_8$ cosmologies by $z=1$.  The comoving number density of halos above $10^{14}h^{-1}M_\odot$ in the $\tau = 50$ Gyr cosmology only increases by a factor of 5 since $z=1$, even though it grows by a factor of $\sim 8$ in the high-$\Omega_m$, high-$\sigma_8$ cosmology, a factor of $\sim 11$ in the mean WMAP-5 cosmology, and a factor of 16 in the low-$\Omega_m$, low-$\sigma_8$ cosmology in the absence of decays.  Thus, a signature of decaying dark matter would be a slower growth of structure with redshift than expected based on the $z=0$ mass function.

A caveat is that a non-cosmological constant ($\Lambda$) model for dark energy could also change the evolution of $n(>M)$ above a fixed mass threshold.  However, it may be possible to break the degeneracy between the two by looking at the evolution of the shape of the cluster mass function.

\subsection{Galaxy clustering}\label{sec:obs:pofk}
The clustering of galaxies is often used to constrain cosmological parameters.  On large, linear scales (which, at $z=0$, corresponds to comoving wavenumbers $k \lesssim 0.1 h\hbox{ Mpc}^{-1}$), clustering is usually expressed in terms of the galaxy power spectrum $P_\mathrm{gg}(k)$, the Fourier transform of the galaxy autocorrelation function.  On smaller, nonlinear scales, the clustering is usually expressed in terms of the real-space galaxy two-point correlation function $\xi(r)$.  Galaxies are biased tracers of the matter distribution, but the matter power spectrum $P(k)$ (or correlation function) can be extracted from galaxy clustering using analytic and empirical models of how galaxies populate dark matter halos.

In this section, I describe how $P(k)$ and $\xi(r)$ could be used to constrain decaying dark matter models, but a detailed calculation is far beyond the scope of this paper.  I will give only a crude, order-of-magnitude limit on decaying dark matter parameter space based on the measured $\xi(r)$ from the Sloan Digital Sky Survey (SDSS) \cite{zehavi2005}.  A proper analysis of the limits on decaying dark matter would require solving the Boltzmann and Einstein equations for the linear matter power spectrum and the collapse criteria for halo formation, as well as $N$-body simulations to better explore the nonlinear regime.

The linear matter power spectrum can be determined from first principles by solving the Einstein equations for first-order perturbations from the background metric and the Boltzmann equations governing the evolution of the distribution functions of all the contents of the Universe.  Calculations typically include CDM, baryons, massive neutrinos, and a cosmological constant $\Lambda$ for the dark energy.  Relative to a $\Lambda$CDM + baryon power spectrum, the $\Lambda$CDM + baryon + neutrino power spectrum is suppressed on scales corresponding the the neutrino free-streaming scale \cite{bond1983}.  This is because neutrinos can escape the matter density perturbations if their typical speed is larger than the escape speed from a perturbation, which reduces the density perturbation.  In decaying dark matter models, the $X$ particles continuously source a free-streaming $Y$-particle populations.  Thus, the density perturbations on small scales (below the typical free-streaming scales of the $Y$ particles) should be damped relative to the $\Lambda$CDM expectation.  However, the power spectrum on large scales should be unaffected by this damping.

The scale-dependence of the damping of the power spectrum due to decays is what may make the effects of decays distinguishable from non-$\Lambda$ dark-energy models.  Unless dark energy clusters, the primary effect of a non-$\Lambda$ dark energy is to alter the growth function, which is scale independent.  Thus, to distinguish between varying-$w$ and decaying dark matter models, one will need to determine the evolution of the shape of the linear matter power spectrum (or the cluster mass function).

On nonlinear scales, the two-point galaxy correlation function $\xi(r)$ is often used to quantify galaxy clustering, instead of the Fourier-space analog preferred on larger scales.  Since galaxies are biased tracers of the underlying matter field, considerable effort has gone into understanding the connection between different galaxy populations and the matter field.  This is also relevant on larger scales, to map between the galaxy power spectrum and the linear matter power spectrum.  The correlation function is often analyzed in the context of halo-occupation distribution (HOD) models, analytic but simulation-calibrated relations describing the distribution $P(N|M)$ of the number $N(M)$ of a galaxies of a certain type contained within a halo of mass $M$ \cite{seljak2000,berlind2002,zheng2002,zheng2005,zheng2007}.  $P(N|M)$ is determined empirically, along with cosmological parameters, for the set of galaxies being analyzed.  There are some degeneracies between the HOD and cosmological parameters, but these can be broken by combining the $\xi(r)$ analysis with different galaxy statistics of the same set of galaxies, or different observations altogether \cite{abazajian2005a,zheng2007,white2009}.
  
HODs make the interpretation of $\xi(r)$ simpler, so I can show why it is difficult to pull easy constraints on decaying dark matter models from the correlation function, and to show what would be necessary in order to use $\xi(r)$ to quantitatively constrain the dark matter model.  The following description closely follows that of Ref. \cite{abazajian2005a}.

The galaxy correlation function can be broken into parts,
\begin{eqnarray}
	\xi(r) = 1 + \xi_{\mathrm{1h}}(r) + \xi_{\mathrm{2h}}(r),
\end{eqnarray}
where $\xi_{\mathrm{1h}}(r)$ is one-halo term, the correlation function of galaxies \emph{within the same halo}, and $\xi_{\mathrm{2h}}(r)$ two-halo term, the correlation function of galaxies in different halos.  The one-halo term is
\begin{multline}
	1 + \xi_{\mathrm{1h}}(r) = \frac{1}{2\pi r^2 \bar{n}^2_g} \int_0^\infty dM \frac{dn}{dM} \frac{\langle N(N-1)\rangle_M}{2} \\
	\times \frac{1}{2R_{\mathrm{vir}}(M)}F^\prime\left( \frac{r}{2R_\mathrm{vir}}\right),
\end{multline}
where $\bar{n}_\mathrm{g}$ is the number density of galaxies, $dn/dM$ is the halo mass function, $\langle N(N-1)\rangle_M/2$ is the average number of pairs of galaxies of a specific type within a halo of mass $M$, and $F^\prime (r/2\Rvire)$ is radial distribution of the galaxy pairs.  $\bar{n}_\mathrm{g}$ is a measured quantity and $\langle N(N-1)\rangle_M/2$ is inferred in the parameter estimation of the correlation function.  However, $dn/dM$ and $F^\prime(r/2\Rvire)$ are taken from dissipationless $N$-body simulations. 

The two-halo term, to lowest order, is given by
\begin{eqnarray}
	\xi_{\mathrm{2h}}( r ) = \frac{1}{2\pi^2} \int_0^\infty dk P^{\mathrm{2h}}_\mathrm{gg}(k) k^2 \frac{\sin kr}{kr},
\end{eqnarray}
where
\begin{multline}
	P^{\mathrm{2h}}_{\mathrm{gg}} = P^{\mathrm{nl}} (k ) \frac{1}{\bar{n}_\mathrm{g}^2} \\
        \times \left[ \int_0^{M_\mathrm{max}} dM \frac{dn}{dM} \langle N(M) \rangle b_\mathrm{h}(M) y_\mathrm{g}(k,M) \right]^2.
\end{multline}
Here, $P^\mathrm{2h}_\mathrm{gg}$ is the two-halo galaxy power spectrum, and $P^\mathrm{nl}$ is the nonlinear matter power spectrum, which is taken from $N$-body simulations.  $\langle N(M) \rangle$, the average number of galaxies in a halo, is determined empirically, but $b_\mathrm{h}$, the bias of the dark matter halo power spectrum with respect to the matter power spectrum, and $y_\mathrm{g}(M,k)$, the Fourier transform of the galaxy density profile in halos, are taken from $N$-body simulations.

The problem with trying to find precision constraints on the decay parameters from $\xi(r)$ is that the interpretation of thereof requires detailed $N$-body simulations to determine $P^{\mathrm{nl}}(k)$, $dn/dM$, $F^\prime(r/2\Rvire)$, $b_\mathrm{g}(M)$, and $y_\mathrm{g}(M,k)$, although a few of those quantities may be found empirically with weak lensing (e.g., Refs \cite{mandelbaum2006,mandelbaum2006a}).  These simulations are generally only performed for ``standard'' cosmologies, and it is not clear how dark matter decays will affect the power spectrum or correlation functions, or interpretation thereof.  Simulations using $\Lambda$CDM cosmologies show that cosmological parameters are somewhat degenerate with the halo occupation models for the galaxies, but the degeneracies have only been explored in a limited set of cosmological models \cite{zheng2002,zheng2007}.

However, there is general consistency of the fits to the correlation function of local SDSS galaxies with $\Lambda$CDM models down to halo masses $\sim 10^{12}M_\odot$ \cite{zehavi2005}.  This suggests that decaying dark matter parameters could be well-constrained using the galaxy correlation function.  In the absence of $N$-body simulations to calibrate HOD models as a function of decaying dark matter parameter space, I constrain the parameter space by requiring that $\sim 10^{12}M_\odot$ $\Lambda$CDM halos lose less than half their mass due to decays.  This roughly restricts $\tau \gtrsim 30$ Gyr for $\vke \gtrsim \vvire \approx 130\hbox{ km s}^{-1}$, which are case-1-type constraints.  It should be noted that this is a highly approximate constraint; more precise constraints require cosmological $N$-body models to calibrate HODs in the case of decaying dark matter.

In addition to simulations to constrain HOD models for decaying dark matter cosmologies, one may constrain the decay parameter space observationally with gravitational lensing.  From gravitational lensing, one may determine the typical halo mass for a specific type of galaxy.  This is an additional constraint to the HOD.  Main halo masses have been found for galaxies binned by stellar mass in the SDSS \cite{mandelbaum2006}, and should be even better-determined in upcoming large all-sky surveys (e.g., the LSST \cite{lsst2009}).

\subsection{Existence of Small ($\sim 10^9M_\odot$) Halos}\label{sec:obs:small}
The smallest virialized halos for which there is observational evidence of existence are of order $10^9M_\odot$.  Halos of such small size are observed in two different ways, and in both cases, the halos are actually subhalos.  Strigari et al. \cite{strigari_nat2008} find that all faint dwarf galaxies in the Milky Way halo have mass interior to 300 pc of $\sim 10^7 M_\odot$ regardless of luminosity.  This corresponds to a virial mass of $\sim 10^9M_\odot$ if the mass profile is extrapolated beyond the stellar component.

Vegetti et al. \cite{vegetti2009c} find evidence for a subhalo of mass $M_{\mathrm{sub}} = (3.51\pm 0.15)\times 10^9 M_\odot$ in the double Einstein ring system SDSSJ0946+1106.  This is the mass within the tidal radius, so the subhalo was likely somewhat larger before it fell into the larger halo of the elliptical galaxy, the lens system.

The existence of these small halos can set limits on a broad swath of decay parameter space.  The most stringent limit is set if one requires that the number density of $10^9M_\odot$-mass halos be the same as predicted by CDM, and that the density profiles within those halos not be significantly disturbed due to decays.  Recalling from Sec. \ref{sec:vir} that the virial speed of such halos is $\sim 13 \hbox{ km s}^{-1}$, one can roughly exclude the decay parameter space above $\vke \sim 10\hbox{ km s}^{-1}$ and $\tau \lesssim$ a few times $t_\mathrm{H}$.  However, there are no $z=0$ observational constraints on the low-mass halo mass function.

As there is no measurement of the number density or correlation function of halos below $\sim 10^{12}M_\odot$, there is considerably more freedom in the decay parameter space.  For example, one could imagine that the $10^9M_\odot$ halos that have been observed at $z=0$ are the remnants of larger halos that have been heavily disturbed by decays.  If today's $\sim 10^9M_\odot$ halos are parented by halos that were initially $10^{10}M_\odot$, then parent halos would have initially had $\vvire \sim 30\hbox{ km s}^{-1}$ and $\tdyne \sim$ a few hundred Myr.  For such halos to lose $\sim 90\%$ of their mass, $\vke \gtrsim \vvire$ and $\tau \ll t_\mathrm{H}$.  This opens up much more allowed decay parameter space than the more previous, more conservative limit.  One could imagine halos of up to $\sim 10^{11}M_\odot$ decaying to $\sim 10^9M_\odot$ halos; in this case, short decay times ($\tau<t_\mathrm{H}$) and large kick speeds ($\vke \gtrsim \vvire \approx 60\hbox{ km s}^{-1}$) are required.  Pushing against this limit is the fact that $\sim 10^{12}M_\odot$ halos appear to have correlation functions consistent with $\Lambda$CDM cosmology.  The virial speed of $10^{11}M_\odot$ halos is a significant fraction of the virial speed of $10^{12}M_\odot$ halos, so such halos would likely be quite disturbed, the degree of which can only be ascertained with simulations of the cases 2 and 4 of Sec. \ref{sec:vir}.

In summary, the existence of $\sim 10^9M_\odot$ halos excludes the parameter space $\vke \gtrsim 60 \hbox{ km s}^{-1}$ and $\tau \lesssim t_\mathrm{H} $.  Stricter constraints come from requiring the mass function of $\sim 10^9-10^{11}M_\odot$ halos to resemble that which is predicted from $\Lambda$CDM models, although I emphasize that there are no observations that require these stricter constraints.  These constraints are similar to those from galaxy correlation functions (Sec. \ref{sec:obs:pofk}).  For $\vke \lesssim 1\hbox{ km s}^{-1}$ or $\tau \gtrsim 10 t_\mathrm{H}$, halos with masses $\gtrsim 10^9M_\odot$ should look and cluster like CDM halos.

\subsection{Mass-Concentration Relation}\label{sec:obs:mc}

$\Lambda$CDM predicts a relationship between the concentration $c$ (Eq. (\ref{eq:rse})) and halo mass, in addition to the shape of the dark matter density profile in halos.  Low-mass halos are expected to be more concentrated than high-mass halos, since smaller halos form earlier when the Universe is more dense.  Concentrations are expected to be higher at the present than at higher redshift for fixed halo mass.  These trends have been found in a number of $N$-body simulations (e.g., Refs. \cite{bullock2001,maccio2007,neto2007,duffy2008,gao2008,maccio2008,rudd2008}) and observations (e.g., Refs. \cite{rines2006,buote2007,comerford2007,mandelbaum2008}), although there is some disagreement in the details.

\begin{figure}
	\includegraphics[width=0.45\textwidth]{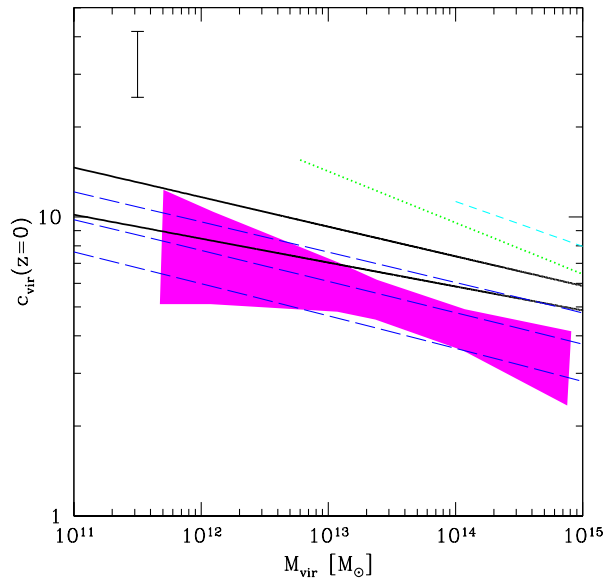}
	\caption{\label{fig:mc}Halo mass-concentration relations at $z=0$. The solid lines are predictions from the $N$-body simulations of Macci{\` o} et al. \cite{maccio2008} for WMAP-1 (upper) and WMAP-3 (lower) cosmologies.  The short-dashed and dotted lines are mass-concentration relations observed in galaxy clusters \cite{comerford2007} and in galaxy groups and clusters \cite{buote2007}.  The symbol in the upper-left corner represents the intrinsic scatter in the simulations and 1-$\sigma$ errors in the mean observed mass-concentration relations.  The shaded region shows the mass-concentration relation in Ref. \cite{mandelbaum2008}.  The long-dashed lines show the mass-concentration relation for high-\vk ~dark matter models after (top to bottom) a fraction $f = 0.1, 0.2$ and $0.3$ of the $X$ particles have decayed.}
\end{figure}

As with the cluster mass function in Sec. \ref{sec:obs:cluster}, I can place constraints on decaying dark matter models by considering the range of mass-concentration relations allowed in WMAP-5 cosmology.  High-$\Omega_m$, high-$\sigma_8$ cosmologies produce higher concentrations for fixed mass than low-$\Omega_m$, low-$\sigma_8$ cosmologies, since halos collapse earlier for fixed mass if the amplitude of the matter power spectrum is higher.  In Fig. \ref{fig:mc}, I show the mean mass-concentration relation for WMAP-1 (upper solid line; $\Omega_m = 0.299$, $\sigma_8 = 0.9$) and WMAP-3 (lower solid line; $\Omega_m = 0.238$, $\sigma_8 = 0.75$) cosmologies from the dark matter-only simulations of Macci{\`o} et al. \cite{maccio2008}.  The values of $\Omega_m$ and $\sigma_8$ bracket the 2-$\sigma$ values of $\Omega_m$ and $\sigma_8$ of the mean WMAP-5 cosmology, and the WMAP-5 mass-concentration relation should lie between the WMAP-1 and WMAP-3 relations \cite{spergel2003,spergel2007,dunkley2009}.  The error bar in the upper-left corner of Fig. \ref{fig:mc} shows the intrinsic scatter in the relation, which is large.

As in Sec. \ref{sec:obs:cluster}, I set conservative constraints on case 1 virial parameters ($\vke > \vvire$, $\tau > t_{\mathrm{dyn}}$, where the halo parameters correspond to the most massive halos observed), by considering the effects of decay on the observables by assuming a high-$\Omega_m$, high-$\sigma_8$ (WMAP-1) cosmology.  I use Eqs. (\ref{eq:rse_new}) and (\ref{eq:rhose_new}) to determine the mass and concentration of a halos after a fraction $f$ of $X$-particles have decayed.  The new mass-concentration relations are shown in Fig. \ref{fig:mc} with long-dashed lines, with (top-to-bottom) $f=0.1,0.2$ and $0.3$.  For $f > 0.2$ ($\tau > 60$ Gyr), the mean mass-concentration relation lies below the WMAP-3 relation from Ref. \cite{maccio2008}.

Unlike Sec. \ref{sec:obs:cluster}, I use the observed mass-concentration relation to constrain the decay parameter space.  I show mass-concentration relations inferred from observations in Fig. \ref{fig:mc}.  The short-dashed line shows the mass-concentration relation found by Ref. \cite{comerford2007} using published masses and concentrations of galaxy clusters.  The masses and concentrations of these clusters were determined using a variety of different observations (e.g., strong lensing, weak lensing, cluster galaxy dynamics, X-ray temperature profiles).  The dotted line shows the mass-concentration relation inferred from X-ray temperature profiles of elliptical galaxies, galaxy groups, and galaxy clusters \cite{buote2007}.  The hatched region shows the 1-$\sigma$ range of mass-concentration relations from weak lensing \cite{mandelbaum2008}.

Of the three observational data sets, the weak lensing data set should be least affected by selection effects.  It has been shown in simulations that strong-lens systems are biased towards high concentrations for fixed mass relative to the population of halos as a whole \cite{hennawi2007,vandeven2009,mandelbaum2009}.  Comerford and Natarajan \cite{comerford2007} show that the mass-concentration relation of strong-lens systems in the simulations of Hennawi et al. \cite{hennawi2007} is a close match to their mass-concentration relation from clusters.  Halo mass profiles can only be reasonably determined from X-ray data if the halo is relaxed; relaxed halos have higher concentration than halos as a whole \cite{neto2007}.  However, selection biases for the X-ray probes of halo properties have not been quantified.

Thus, I compare the mass-concentration relation from the decaying dark matter models to the weak-lensing relation.  I find that for $f=0.3$ ($\tau = 40$ Gyr), the decaying dark matter mass-concentration relation lies well below the mean relation from weak lensing \cite{mandelbaum2008}.  This sets the lower limit on the allowed value of $\tau$ for high \vk, where $\vke > \vvire$ for the largest clusters probed in the observations ($\vvire \gtrsim$ a few $\times 1000\hbox{ km s}^{-1}$).  This limit is nearly identical to that obtained from the cluster mass function in Sec. \ref{sec:obs:cluster}.

\section{Discussion}\label{sec:conclusion}
\subsection{Summary}\label{sec:conc:sum}

\begin{figure}[t]
	\includegraphics[width=0.4\textwidth,angle=270]{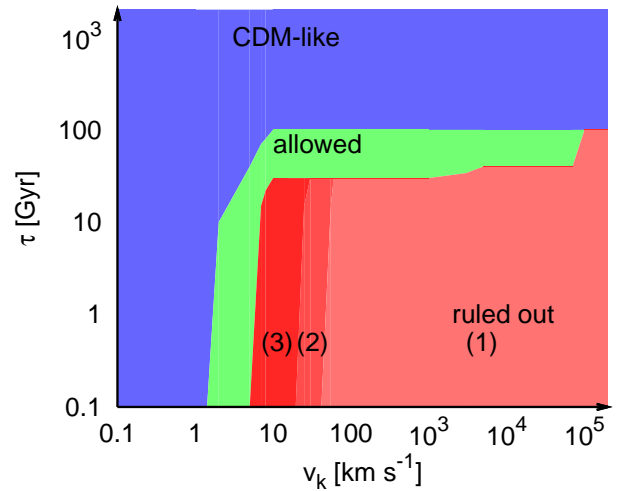}
	\caption{\label{fig:phys}Diagram of allowed and excluded $\vke-\tau$ parameter space.  See text for descriptions.}
\end{figure}

The constraints are summarized in Fig. \ref{fig:phys}.  At near-relativistic to relativistic \vk, the only allowed region of parameter space is $\tau > 120$ Gyr, which is nearly indistinguishable from CDM, as only $\sim$10\% of $X$ particles will have decayed by the present.  The region with $\tau \lesssim 20$ Gyr yields catastrophic destruction of halos on all scales, while $20 \hbox{ Gyr} < \tau < 120 \hbox{ Gyr}$ is ruled out by the shape of the CMB temperature power spectrum \cite{ichiki2004}, although additional constraints may be possible using luminosity distances of supernovae \cite{zentner2002}.

For kick speeds of order $\gtrsim 5000\hbox{ km s}^{-1}$, both the cluster mass functions and the mass-concentration relation restricts $\tau \gtrsim 40$ Gyr.  A calculation of the cosmological Boltzmann and Einstein equations including decay to find the linear matter power spectrum, and $N$-body simulations to explore the nonlinear regime, should yield even better constraints on $\tau$ in this \vk~regime.

For $60\hbox{ km s}^{-1} \lesssim \vke \lesssim 5000 \hbox{ km s}^{-1}$, $\tau \gtrsim 30$ Gyr based on $\xi(r)$ and the existence of $\sim 10^9M_\odot$ halos.  The excluded region is marked (1) in Fig. \ref{fig:phys}.  Better constraints on $\tau$ for this range of \vk~ using $\xi(r)$ will come from cosmological simulations of structure formation that include decay.  It may be easier to constrain this part of decay parameter space using mass-concentration relations or the cluster mass function.  In this case, one can do simulations of isolated halos in the regimes $\tau < \tdyne$ and $\vke < \vvire$ \& $\tau > \tdyne$ to map the $\Lambda$CDM mass-concentration relation and cluster mass function to those with cosmologies with decaying dark matter.  We perform such simulations and will present our findings in another paper \cite{peter2010}.

Any $\tau$ is allowed for $1\hbox{ km s}^{-1} \lesssim \vke \lesssim 60\hbox{ km s}^{-1}$ unless one uses the more conservative constraints on the parameter space based on the existence of a few $\sim 10^9M_\odot$ halos.  The additional exclusion zones are marked (2) \& (3) in Fig. \ref{fig:phys}, corresponding to allowing $10^9M_\odot$ halos to result from decays of halos that initially had mass $10^{10}M_\odot$ and from requiring that the halo number density $n(>10^9M_\odot)$ be nearly indistinguishable from $\Lambda$CDM, respectively.  I emphasize that there are no observations that definitively exclude these additional regions of parameter space.  In general, these regions of parameter space can be better constrained theoretically by the aforementioned cosmological simulations, in order to relate the observed $\xi(r)$ to decaying dark matter parameter space.  It may be possible to constrain some of this parameter space using the mass-concentration relation, as weak lensing has already probed this relation down to halo masses $M_\mathrm{vir} < 10^{12}M_\odot$.  Future deep surveys covering a large fraction of the sky (e.g, LSST \cite{lsst2009}) should improve and extend measurements of the mass-concentration relation to lower halo masses.

For $\vke < 1\hbox{ km s}^{-1}$, there will be negligible deviations from $\Lambda$CDM cosmology on $\sim 10^9M_\odot$ halo scales and up, and negligible differences on all scales if $\tau \gtrsim 100$ Gyr, for the simple reason that structure does not look significantly different from $\Lambda$CDM if almost no particles decay or if the daughter particles receive a kick small relative to the virial speed.  This part of parameter space is labeled ``CDM-like'' in Fig. \ref{fig:phys} for those regions.  If halos smaller than $10^9M_\odot$ are observed, this would push the ``CDM-like'' line at $\vke < 1\hbox{ km s}^{-1}$ down to smaller \vk.  The discovery of smaller (sub)halos is likely in future observations of strongly lensed galaxies, since flux anomalies and time delays are sensitive to the subhalo mass function down to near-stellar mass scales \cite{dalal2002,keeton2009,koopmans2009,marshall2009,moustakas2009}.  The absence of smaller subhalos would have profound implications for the nature of dark matter.

\subsection{Why I am not using the dark matter density profile as a constraint}\label{sec:obs:nfw}

Dissipationless $N$-body simulations of structure formation suggest that the density profiles of dark matter halos should be well described by the NFW profile of Eq. (\ref{eq:nfw}) on any observable scale.  Although decays in the case-1 regime to not alter the structure of dark matter halos, decays in other regimes generically change the halo structure due to kinetic-energy injection.  Thus, in principle, the density profile in dark matter halos would appear to be a good test of the decaying dark matter model.

However, observations of galaxies and clusters, from the smallest ultrafaint dwarfs in the Local Group to the largest virialized halos in the local Universe, show significant scatter in the density profile on scales $r < \rse$.  Some observations indicate cores in the density profile, others show cusps shallower than NFW, some show that the NFW profile is a reasonable fit, and others show profiles that are more sharply cusped than NFW (for a summary, see Refs. \cite{gilmore2007,strigari2007d,strigari2008,walker2009} on Milky Way dwarfs, \cite{simon2005,kuzio2006,bailin2007} on the rotation curves of low-surface-brightness galaxies, \cite{deblok2008,oh2008,trott2009} on the rotation curves of massive spiral galaxies, \cite{treu2004,padmanabhan2004,dekel2005,humphrey2006,mandelbaum2006,schulz2009} on lensing and kinematics of elliptical galaxies, and \cite{comerford2006,mandelbaum2006a,rines2006,gastaldello2007,limousin2007,sand2008,oguri2009,newman2009} on various probes of the dark matter distribution in galaxy groups and clusters).

While there are likely to be some observational systematics that affect the dark matter profile fits, perhaps the most significant issue in interpreting these data is that there are serious theoretical systematics.  The NFW profile emerges in simulations of structure formation without baryons.  However, galaxies dominate the gravitational potential in the inner parts of the dark matter halos (with the possible exceptions of the ultrafaint dwarfs in the Local Group and some low-surface-brightness galaxies).  There is evidence from simulations that the dark matter responds to various processes associated with galaxy evolution (feedback, star formation rate, gas cooling, smooth or clumpy baryonic infall), but it is not clear which processes dominate \cite{elzant2001,romano-diaz2008,abadi2009,pedrosa2010,pedrosa2009a,romano-diaz2009,tissera2009}.  Dark matter halos in $N$-body simulations with CDM and baryons range from having cored density profiles \cite{governato2009} to quite cusped profiles \cite{gnedin2004,gustafsson2006}.  At this point, the interpretation of simulations is descriptive rather than predictive.  What is needed in order to interpret observations are concrete theoretical predictions, not only for CDM, but for other types of dark matter. 

Thus, even though dark matter decay is likely to change the density profile of dark matter in halos from the form of Eq. (\ref{eq:nfw}), the observed density profiles will not constrain decaying dark matter parameter space until there are firmer theoretical predictions for the dark matter distribution in halos in the presence of a baryonic galaxy.  This is especially important because decays affect the central regions of halos, the home of galaxies, more strongly than the outside regions for fixed $\tau$ and \vk, since the dynamical time scales and typical particle velocities are lower.

\subsection{Future directions}\label{sec:conc:future}
The easiest way to constrain more of the decay parameter space is with the mass-concentration relation and cluster mass functions.  Future observational data will improve constraints on the decay parameter space.  Upcoming deep optical all-sky surveys such as DES \cite{annis2005}, PanSTARRS \cite{kaiser2002}, and LSST \cite{lsst2009} will allow for a better determination of the mass-concentration relation via weak lensing.  The redshift-dependent cluster mass function may be determined with next-generation optical all-sky surveys, as well as Sunyaev-Zel'dovich surveys (SPT \cite{staniszewski2009}; ACT \cite{hincks2009}) and X-ray surveys (eROSITA \cite{predehl2006}).  These data sets, along with CMB measurements with \emph{Planck}, will provide better constraints on cosmological parameters, which will further constrain the decay parameter space.  The effects of decay on the mass-concentration relation and cluster mass functions is relatively easy to quantify since these constraints do not require cosmological simulations.  In this work, I showed how to constrain $\tau$ in the case that \vk~is significantly larger than the virial speed of the largest galaxy cluster.  To constrain other parts of the decay parameter space, one may do noncosmological simulations of isolated dark matter halos.  This will allow one to determine the change to the halo structure and mass loss as a function of decay parameters, and is investigated in Ref. \cite{peter2010}.

\begin{acknowledgments}
I thank Marc Kamionkowski and Chris Hirata for useful conversations, and James Bullock and Jeremy Tinker for their mass-concentration and halo mass-function codes.  This work was supported by the Gordon and Betty Moore Foundation.
\end{acknowledgments}

%\bibliographystyle{elsarticle-num}
%\bibliography{dmrefs}

\begin{thebibliography}{136}
\expandafter\ifx\csname natexlab\endcsname\relax\def\natexlab#1{#1}\fi
\expandafter\ifx\csname bibnamefont\endcsname\relax
  \def\bibnamefont#1{#1}\fi
\expandafter\ifx\csname bibfnamefont\endcsname\relax
  \def\bibfnamefont#1{#1}\fi
\expandafter\ifx\csname citenamefont\endcsname\relax
  \def\citenamefont#1{#1}\fi
\expandafter\ifx\csname url\endcsname\relax
  \def\url#1{\texttt{#1}}\fi
\expandafter\ifx\csname urlprefix\endcsname\relax\def\urlprefix{URL }\fi
\providecommand{\bibinfo}[2]{#2}
\providecommand{\eprint}[2][]{\url{#2}}

\bibitem[{\citenamefont{{Komatsu} et~al.}(2009)}]{komatsu2009}
\bibinfo{author}{\bibfnamefont{E.}~\bibnamefont{{Komatsu}}}
  \bibnamefont{et~al.}, \bibinfo{journal}{Astrophys. J. Suppl. Ser.}
  \textbf{\bibinfo{volume}{180}}, \bibinfo{pages}{330} (\bibinfo{year}{2009}).

\bibitem[{\citenamefont{{Frieman} et~al.}(2008)\citenamefont{{Frieman},
  {Turner}, and {Huterer}}}]{frieman2008}
\bibinfo{author}{\bibfnamefont{J.~A.} \bibnamefont{{Frieman}}},
  \bibinfo{author}{\bibfnamefont{M.~S.} \bibnamefont{{Turner}}},
  \bibnamefont{and}
  \bibinfo{author}{\bibfnamefont{D.}~\bibnamefont{{Huterer}}},
  \bibinfo{journal}{Annu. Rev. Astron. Astrophs.} \textbf{\bibinfo{volume}{46}},
  \bibinfo{pages}{385} (\bibinfo{year}{2008}).

\bibitem[{\citenamefont{{Caldwell} and {Kamionkowski}}(2009)}]{caldwell2009}
\bibinfo{author}{\bibfnamefont{R.~R.} \bibnamefont{{Caldwell}}}
  \bibnamefont{and}
  \bibinfo{author}{\bibfnamefont{M.}~\bibnamefont{{Kamionkowski}}},
  \bibinfo{journal}{Annu. Rev. Nucl. Part. Sci.}
  \textbf{\bibinfo{volume}{59}}, \bibinfo{pages}{397} (\bibinfo{year}{2009}).

\bibitem[{\citenamefont{{Jungman} et~al.}(1996)\citenamefont{{Jungman},
  {Kamionkowski}, and {Griest}}}]{jungman1996}
\bibinfo{author}{\bibfnamefont{G.}~\bibnamefont{{Jungman}}},
  \bibinfo{author}{\bibfnamefont{M.}~\bibnamefont{{Kamionkowski}}},
  \bibnamefont{and} \bibinfo{author}{\bibfnamefont{K.}~\bibnamefont{{Griest}}},
  \bibinfo{journal}{Phys. Rep.} \textbf{\bibinfo{volume}{267}},
  \bibinfo{pages}{195} (\bibinfo{year}{1996}).

\bibitem[{\citenamefont{{Cheng} et~al.}(2002)\citenamefont{{Cheng}, {Feng}, and
  {Matchev}}}]{cheng2002}
\bibinfo{author}{\bibfnamefont{H.-C.} \bibnamefont{{Cheng}}},
  \bibinfo{author}{\bibfnamefont{J.~L.} \bibnamefont{{Feng}}},
  \bibnamefont{and} \bibinfo{author}{\bibfnamefont{K.~T.}
  \bibnamefont{{Matchev}}}, \bibinfo{journal}{Phys. Rev. Lett.}
  \textbf{\bibinfo{volume}{89}}, \bibinfo{pages}{211301}
  (\bibinfo{year}{2002}).

\bibitem[{\citenamefont{{Tegmark} et~al.}(2004)\citenamefont{{Tegmark},
  {Strauss}, {Blanton}, {Abazajian}, {Dodelson}, {Sandvik}, {Wang}, {Weinberg},
  {Zehavi}, {Bahcall} et~al.}}]{tegmark2004}
\bibinfo{author}{\bibfnamefont{M.}~\bibnamefont{{Tegmark}}},
  \bibinfo{author}{\bibfnamefont{M.~A.} \bibnamefont{{Strauss}}},
  \bibinfo{author}{\bibfnamefont{M.~R.} \bibnamefont{{Blanton}}},
  \bibinfo{author}{\bibfnamefont{K.}~\bibnamefont{{Abazajian}}},
  \bibinfo{author}{\bibfnamefont{S.}~\bibnamefont{{Dodelson}}},
  \bibinfo{author}{\bibfnamefont{H.}~\bibnamefont{{Sandvik}}},
  \bibinfo{author}{\bibfnamefont{X.}~\bibnamefont{{Wang}}},
  \bibinfo{author}{\bibfnamefont{D.~H.} \bibnamefont{{Weinberg}}},
  \bibinfo{author}{\bibfnamefont{I.}~\bibnamefont{{Zehavi}}},
  \bibinfo{author}{\bibfnamefont{N.~A.} \bibnamefont{{Bahcall}}}
  \bibnamefont{et~al.}, \bibinfo{journal}{\prd} \textbf{\bibinfo{volume}{69}},
  \bibinfo{pages}{103501} (\bibinfo{year}{2004}).

\bibitem[{\citenamefont{{Lesgourgues} et~al.}(2007)\citenamefont{{Lesgourgues},
  {Viel}, {Haehnelt}, and {Massey}}}]{lesgourgues2007}
\bibinfo{author}{\bibfnamefont{J.}~\bibnamefont{{Lesgourgues}}},
  \bibinfo{author}{\bibfnamefont{M.}~\bibnamefont{{Viel}}},
  \bibinfo{author}{\bibfnamefont{M.~G.} \bibnamefont{{Haehnelt}}},
  \bibnamefont{and} \bibinfo{author}{\bibfnamefont{R.}~\bibnamefont{{Massey}}},
  \bibinfo{journal}{J. Cosmol. Astropart. Phys.}
  \textbf{\bibinfo{volume}{11}}, \bibinfo{pages}{008} (\bibinfo{year}{2007}).

\bibitem[{\citenamefont{{Reid} et~al.}(2009)}]{breid2009}
\bibinfo{author}{\bibfnamefont{B.~A.} \bibnamefont{{Reid}}}
  \bibnamefont{et~al.}, \bibinfo{journal}{arXiv:0907.1659}.

\bibitem[{\citenamefont{{Vikhlinin}
  et~al.}(2009{\natexlab{a}})\citenamefont{{Vikhlinin}, {Kravtsov}, {Burenin},
  {Ebeling}, {Forman}, {Hornstrup}, {Jones}, {Murray}, {Nagai}, {Quintana}
  et~al.}}]{vikhlinin2009b}
\bibinfo{author}{\bibfnamefont{A.}~\bibnamefont{{Vikhlinin}}},
  \bibinfo{author}{\bibfnamefont{A.~V.} \bibnamefont{{Kravtsov}}},
  \bibinfo{author}{\bibfnamefont{R.~A.} \bibnamefont{{Burenin}}},
  \bibinfo{author}{\bibfnamefont{H.}~\bibnamefont{{Ebeling}}},
  \bibinfo{author}{\bibfnamefont{W.~R.} \bibnamefont{{Forman}}},
  \bibinfo{author}{\bibfnamefont{A.}~\bibnamefont{{Hornstrup}}},
  \bibinfo{author}{\bibfnamefont{C.}~\bibnamefont{{Jones}}},
  \bibinfo{author}{\bibfnamefont{S.~S.} \bibnamefont{{Murray}}},
  \bibinfo{author}{\bibfnamefont{D.}~\bibnamefont{{Nagai}}},
  \bibinfo{author}{\bibfnamefont{H.}~\bibnamefont{{Quintana}}}
  \bibnamefont{et~al.}, \bibinfo{journal}{\apj} \textbf{\bibinfo{volume}{692}},
  \bibinfo{pages}{1060} (\bibinfo{year}{2009}{\natexlab{a}}).

\bibitem[{\citenamefont{{Fillmore} and {Goldreich}}(1984)}]{fillmore1984}
\bibinfo{author}{\bibfnamefont{J.~A.} \bibnamefont{{Fillmore}}}
  \bibnamefont{and}
  \bibinfo{author}{\bibfnamefont{P.}~\bibnamefont{{Goldreich}}},
  \bibinfo{journal}{Astrophys. J.} \textbf{\bibinfo{volume}{281}},
  \bibinfo{pages}{1} (\bibinfo{year}{1984}).

\bibitem[{\citenamefont{{Hofmann} et~al.}(2001)\citenamefont{{Hofmann},
  {Schwarz}, and {St{\"o}cker}}}]{hofmann2001}
\bibinfo{author}{\bibfnamefont{S.}~\bibnamefont{{Hofmann}}},
  \bibinfo{author}{\bibfnamefont{D.~J.} \bibnamefont{{Schwarz}}},
  \bibnamefont{and}
  \bibinfo{author}{\bibfnamefont{H.}~\bibnamefont{{St{\"o}cker}}},
  \bibinfo{journal}{Phys. Rev. D} \textbf{\bibinfo{volume}{64}},
  \bibinfo{pages}{083507} (\bibinfo{year}{2001}).

\bibitem[{\citenamefont{{Profumo} et~al.}(2006)\citenamefont{{Profumo},
  {Sigurdson}, and {Kamionkowski}}}]{profumo2006}
\bibinfo{author}{\bibfnamefont{S.}~\bibnamefont{{Profumo}}},
  \bibinfo{author}{\bibfnamefont{K.}~\bibnamefont{{Sigurdson}}},
  \bibnamefont{and}
  \bibinfo{author}{\bibfnamefont{M.}~\bibnamefont{{Kamionkowski}}},
  \bibinfo{journal}{Phys. Rev. Lett.} \textbf{\bibinfo{volume}{97}},
  \bibinfo{pages}{031301} (\bibinfo{year}{2006}).

\bibitem[{\citenamefont{{Navarro} et~al.}(2010)\citenamefont{{Navarro},
  {Ludlow}, {Springel}, {Wang}, {Vogelsberger}, {White}, {Jenkins}, {Frenk},
  and {Helmi}}}]{navarro2009}
\bibinfo{author}{\bibfnamefont{J.~F.} \bibnamefont{{Navarro}}},
  \bibinfo{author}{\bibfnamefont{A.}~\bibnamefont{{Ludlow}}},
  \bibinfo{author}{\bibfnamefont{V.}~\bibnamefont{{Springel}}},
  \bibinfo{author}{\bibfnamefont{J.}~\bibnamefont{{Wang}}},
  \bibinfo{author}{\bibfnamefont{M.}~\bibnamefont{{Vogelsberger}}},
  \bibinfo{author}{\bibfnamefont{S.~D.~M.} \bibnamefont{{White}}},
  \bibinfo{author}{\bibfnamefont{A.}~\bibnamefont{{Jenkins}}},
  \bibinfo{author}{\bibfnamefont{C.~S.} \bibnamefont{{Frenk}}},
  \bibnamefont{and} \bibinfo{author}{\bibfnamefont{A.}~\bibnamefont{{Helmi}}},
  \bibinfo{journal}{Mon. Not. R. Astron. Soc.} 402, \bibinfo{pages}{21}
  (\bibinfo{year}{2010}).

\bibitem[{\citenamefont{{Dalcanton} and {Hogan}}(2001)}]{dalcanton2001}
\bibinfo{author}{\bibfnamefont{J.~J.} \bibnamefont{{Dalcanton}}}
  \bibnamefont{and} \bibinfo{author}{\bibfnamefont{C.~J.}
  \bibnamefont{{Hogan}}}, \bibinfo{journal}{Astrophys. J.}
  \textbf{\bibinfo{volume}{561}}, \bibinfo{pages}{35} (\bibinfo{year}{2001}).

\bibitem[{\citenamefont{{Simon} et~al.}(2005)\citenamefont{{Simon}, {Bolatto},
  {Leroy}, {Blitz}, and {Gates}}}]{simon2005}
\bibinfo{author}{\bibfnamefont{J.~D.} \bibnamefont{{Simon}}},
  \bibinfo{author}{\bibfnamefont{A.~D.} \bibnamefont{{Bolatto}}},
  \bibinfo{author}{\bibfnamefont{A.}~\bibnamefont{{Leroy}}},
  \bibinfo{author}{\bibfnamefont{L.}~\bibnamefont{{Blitz}}}, \bibnamefont{and}
  \bibinfo{author}{\bibfnamefont{E.~L.} \bibnamefont{{Gates}}},
  \bibinfo{journal}{Astrophys. J.} \textbf{\bibinfo{volume}{621}},
  \bibinfo{pages}{757} (\bibinfo{year}{2005}).

\bibitem[{\citenamefont{{Spergel} and {Steinhardt}}(2000)}]{spergel2000}
\bibinfo{author}{\bibfnamefont{D.~N.} \bibnamefont{{Spergel}}}
  \bibnamefont{and} \bibinfo{author}{\bibfnamefont{P.~J.}
  \bibnamefont{{Steinhardt}}}, \bibinfo{journal}{Phys. Rev. Lett.}
  \textbf{\bibinfo{volume}{84}}, \bibinfo{pages}{3760} (\bibinfo{year}{2000}).

\bibitem[{\citenamefont{{Feng} et~al.}(2003{\natexlab{a}})\citenamefont{{Feng},
  {Rajaraman}, and {Takayama}}}]{feng2003}
\bibinfo{author}{\bibfnamefont{J.~L.} \bibnamefont{{Feng}}},
  \bibinfo{author}{\bibfnamefont{A.}~\bibnamefont{{Rajaraman}}},
  \bibnamefont{and}
  \bibinfo{author}{\bibfnamefont{F.}~\bibnamefont{{Takayama}}},
  \bibinfo{journal}{Phys. Rev. Lett.} \textbf{\bibinfo{volume}{91}},
  \bibinfo{pages}{011302} (\bibinfo{year}{2003}{\natexlab{a}}).

\bibitem[{\citenamefont{{Cembranos} et~al.}(2005)\citenamefont{{Cembranos},
  {Feng}, {Rajaraman}, and {Takayama}}}]{cembranos2005}
\bibinfo{author}{\bibfnamefont{J.~A.} \bibnamefont{{Cembranos}}},
  \bibinfo{author}{\bibfnamefont{J.~L.} \bibnamefont{{Feng}}},
  \bibinfo{author}{\bibfnamefont{A.}~\bibnamefont{{Rajaraman}}},
  \bibnamefont{and}
  \bibinfo{author}{\bibfnamefont{F.}~\bibnamefont{{Takayama}}},
  \bibinfo{journal}{Phys. Rev. Lett.} \textbf{\bibinfo{volume}{95}},
  \bibinfo{pages}{181301} (\bibinfo{year}{2005}).

\bibitem[{\citenamefont{{Profumo} et~al.}(2005)\citenamefont{{Profumo},
  {Sigurdson}, {Ullio}, and {Kamionkowski}}}]{profumo2005}
\bibinfo{author}{\bibfnamefont{S.}~\bibnamefont{{Profumo}}},
  \bibinfo{author}{\bibfnamefont{K.}~\bibnamefont{{Sigurdson}}},
  \bibinfo{author}{\bibfnamefont{P.}~\bibnamefont{{Ullio}}}, \bibnamefont{and}
  \bibinfo{author}{\bibfnamefont{M.}~\bibnamefont{{Kamionkowski}}},
  \bibinfo{journal}{\prd} \textbf{\bibinfo{volume}{71}},
  \bibinfo{pages}{023518} (\bibinfo{year}{2005}).

\bibitem[{\citenamefont{{S{\'a}nchez-Salcedo}}(2003)}]{sanchez2003}
\bibinfo{author}{\bibfnamefont{F.~J.} \bibnamefont{{S{\'a}nchez-Salcedo}}},
  \bibinfo{journal}{\apj} \textbf{\bibinfo{volume}{591}}, \bibinfo{pages}{L107}
  (\bibinfo{year}{2003}).

\bibitem[{\citenamefont{{Feng} and {Kumar}}(2008)}]{feng2008}
\bibinfo{author}{\bibfnamefont{J.~L.} \bibnamefont{{Feng}}} \bibnamefont{and}
  \bibinfo{author}{\bibfnamefont{J.}~\bibnamefont{{Kumar}}},
  \bibinfo{journal}{Phys. Rev. Lett.} \textbf{\bibinfo{volume}{101}},
  \bibinfo{pages}{231301} (\bibinfo{year}{2008}).

\bibitem[{\citenamefont{{Feng} et~al.}(2008)\citenamefont{{Feng}, {Tu}, and
  {Yu}}}]{feng2008a}
\bibinfo{author}{\bibfnamefont{J.~L.} \bibnamefont{{Feng}}},
  \bibinfo{author}{\bibfnamefont{H.}~\bibnamefont{{Tu}}}, \bibnamefont{and}
  \bibinfo{author}{\bibfnamefont{H.-B.} \bibnamefont{{Yu}}},
  \bibinfo{journal}{J. Cosmol. Astropart. Phys.}
  \textbf{\bibinfo{volume}{10}}, \bibinfo{pages}{43} (\bibinfo{year}{2008}).

\bibitem[{\citenamefont{{Feng} et~al.}(2009)\citenamefont{{Feng}, {Kaplinghat},
  {Tu}, and {Yu}}}]{feng2009}
\bibinfo{author}{\bibfnamefont{J.~L.} \bibnamefont{{Feng}}},
  \bibinfo{author}{\bibfnamefont{M.}~\bibnamefont{{Kaplinghat}}},
  \bibinfo{author}{\bibfnamefont{H.}~\bibnamefont{{Tu}}}, \bibnamefont{and}
  \bibinfo{author}{\bibfnamefont{H.-B.} \bibnamefont{{Yu}}},
  \bibinfo{journal}{J. Cosmol. Astropart. Phys.}
  \textbf{\bibinfo{volume}{07}}, \bibinfo{pages}{004} (\bibinfo{year}{2009}).

\bibitem[{\citenamefont{{Ackerman} et~al.}(2009)\citenamefont{{Ackerman},
  {Buckley}, {Carroll}, and {Kamionkowski}}}]{ackerman2009}
\bibinfo{author}{\bibfnamefont{L.}~\bibnamefont{{Ackerman}}},
  \bibinfo{author}{\bibfnamefont{M.~R.} \bibnamefont{{Buckley}}},
  \bibinfo{author}{\bibfnamefont{S.~M.} \bibnamefont{{Carroll}}},
  \bibnamefont{and}
  \bibinfo{author}{\bibfnamefont{M.}~\bibnamefont{{Kamionkowski}}},
  \bibinfo{journal}{\prd} \textbf{\bibinfo{volume}{79}},
  \bibinfo{pages}{023519} (\bibinfo{year}{2009}).

\bibitem[{\citenamefont{{Doroshkevich} and {Khlopov}}(1984)}]{doroshkevich1984}
\bibinfo{author}{\bibfnamefont{A.~G.} \bibnamefont{{Doroshkevich}}}
  \bibnamefont{and} \bibinfo{author}{\bibfnamefont{M.~Iu.}
  \bibnamefont{{Khlopov}}}, \bibinfo{journal}{Mon. Not. R. Astron. Soc.}
  \textbf{\bibinfo{volume}{211}}, \bibinfo{pages}{277} (\bibinfo{year}{1984}).

\bibitem[{\citenamefont{{Ellis} et~al.}(1985)\citenamefont{{Ellis},
  {Nanopoulos}, and {Sarkar}}}]{ellis1985}
\bibinfo{author}{\bibfnamefont{J.}~\bibnamefont{{Ellis}}},
  \bibinfo{author}{\bibfnamefont{D.~V.} \bibnamefont{{Nanopoulos}}},
  \bibnamefont{and} \bibinfo{author}{\bibfnamefont{S.}~\bibnamefont{{Sarkar}}},
  \bibinfo{journal}{Nucl. Phys. B} \textbf{\bibinfo{volume}{259}},
  \bibinfo{pages}{175} (\bibinfo{year}{1985}).

\bibitem[{\citenamefont{{Kawasaki} and {Moroi}}(1995)}]{kawasaki1995}
\bibinfo{author}{\bibfnamefont{M.}~\bibnamefont{{Kawasaki}}} \bibnamefont{and}
  \bibinfo{author}{\bibfnamefont{T.}~\bibnamefont{{Moroi}}},
  \bibinfo{journal}{\apj} \textbf{\bibinfo{volume}{452}}, \bibinfo{pages}{506}
  (\bibinfo{year}{1995}).

\bibitem[{\citenamefont{{Feng} et~al.}(2003{\natexlab{b}})\citenamefont{{Feng},
  {Rajaraman}, and {Takayama}}}]{feng2003a}
\bibinfo{author}{\bibfnamefont{J.~L.} \bibnamefont{{Feng}}},
  \bibinfo{author}{\bibfnamefont{A.}~\bibnamefont{{Rajaraman}}},
  \bibnamefont{and}
  \bibinfo{author}{\bibfnamefont{F.}~\bibnamefont{{Takayama}}},
  \bibinfo{journal}{\prd} \textbf{\bibinfo{volume}{68}},
  \bibinfo{pages}{063504} (\bibinfo{year}{2003}{\natexlab{b}}).

\bibitem[{\citenamefont{{Chen} and {Kamionkowski}}(2004)}]{chen2004}
\bibinfo{author}{\bibfnamefont{X.}~\bibnamefont{{Chen}}} \bibnamefont{and}
  \bibinfo{author}{\bibfnamefont{M.}~\bibnamefont{{Kamionkowski}}},
  \bibinfo{journal}{\prd} \textbf{\bibinfo{volume}{70}},
  \bibinfo{pages}{043502} (\bibinfo{year}{2004}).

\bibitem[{\citenamefont{{Feng} et~al.}(2004)\citenamefont{{Feng}, {Su}, and
  {Takayama}}}]{feng2004}
\bibinfo{author}{\bibfnamefont{J.~L.} \bibnamefont{{Feng}}},
  \bibinfo{author}{\bibfnamefont{S.}~\bibnamefont{{Su}}}, \bibnamefont{and}
  \bibinfo{author}{\bibfnamefont{F.}~\bibnamefont{{Takayama}}},
  \bibinfo{journal}{\prd} \textbf{\bibinfo{volume}{70}},
  \bibinfo{pages}{063514} (\bibinfo{year}{2004}).

\bibitem[{\citenamefont{{Kaplinghat}}(2005)}]{kaplinghat2005}
\bibinfo{author}{\bibfnamefont{M.}~\bibnamefont{{Kaplinghat}}},
  \bibinfo{journal}{\prd} \textbf{\bibinfo{volume}{72}},
  \bibinfo{pages}{063510} (\bibinfo{year}{2005}).

\bibitem[{\citenamefont{{Cembranos} et~al.}(2007)\citenamefont{{Cembranos},
  {Feng}, and {Strigari}}}]{cembranos2007}
\bibinfo{author}{\bibfnamefont{J.~A.~R.} \bibnamefont{{Cembranos}}},
  \bibinfo{author}{\bibfnamefont{J.~L.} \bibnamefont{{Feng}}},
  \bibnamefont{and} \bibinfo{author}{\bibfnamefont{L.~E.}
  \bibnamefont{{Strigari}}}, \bibinfo{journal}{Phys. Rev. Lett.}
  \textbf{\bibinfo{volume}{99}}, \bibinfo{pages}{191301}
  (\bibinfo{year}{2007}).

\bibitem[{\citenamefont{{Cembranos} and {Strigari}}(2008)}]{cembranos2008}
\bibinfo{author}{\bibfnamefont{J.~A.~R.} \bibnamefont{{Cembranos}}}
  \bibnamefont{and} \bibinfo{author}{\bibfnamefont{L.~E.}
  \bibnamefont{{Strigari}}}, \bibinfo{journal}{\prd}
  \textbf{\bibinfo{volume}{77}}, \bibinfo{pages}{123519}
  (\bibinfo{year}{2008}).

\bibitem[{\citenamefont{{Borzumati} et~al.}(2008)\citenamefont{{Borzumati},
  {Bringmann}, and {Ullio}}}]{borzumati2008}
\bibinfo{author}{\bibfnamefont{F.}~\bibnamefont{{Borzumati}}},
  \bibinfo{author}{\bibfnamefont{T.}~\bibnamefont{{Bringmann}}},
  \bibnamefont{and} \bibinfo{author}{\bibfnamefont{P.}~\bibnamefont{{Ullio}}},
  \bibinfo{journal}{\prd} \textbf{\bibinfo{volume}{77}},
  \bibinfo{pages}{063514} (\bibinfo{year}{2008}).

\bibitem[{\citenamefont{{Bryan} and {Norman}}(1998)}]{bryan1998}
\bibinfo{author}{\bibfnamefont{G.~L.} \bibnamefont{{Bryan}}} \bibnamefont{and}
  \bibinfo{author}{\bibfnamefont{M.~L.} \bibnamefont{{Norman}}},
  \bibinfo{journal}{\apj} \textbf{\bibinfo{volume}{495}}, \bibinfo{pages}{80}
  (\bibinfo{year}{1998}).

\bibitem[{\citenamefont{{Navarro} et~al.}(1997)\citenamefont{{Navarro},
  {Frenk}, and {White}}}]{navarro1997}
\bibinfo{author}{\bibfnamefont{J.~F.} \bibnamefont{{Navarro}}},
  \bibinfo{author}{\bibfnamefont{C.~S.} \bibnamefont{{Frenk}}},
  \bibnamefont{and} \bibinfo{author}{\bibfnamefont{S.~D.~M.}
  \bibnamefont{{White}}}, \bibinfo{journal}{\apj}
  \textbf{\bibinfo{volume}{490}}, \bibinfo{pages}{493} (\bibinfo{year}{1997}).

\bibitem[{\citenamefont{{Navarro} et~al.}(2004)\citenamefont{{Navarro},
  {Hayashi}, {Power}, {Jenkins}, {Frenk}, {White}, {Springel}, {Stadel}, and
  {Quinn}}}]{navarro2004}
\bibinfo{author}{\bibfnamefont{J.~F.} \bibnamefont{{Navarro}}},
  \bibinfo{author}{\bibfnamefont{E.}~\bibnamefont{{Hayashi}}},
  \bibinfo{author}{\bibfnamefont{C.}~\bibnamefont{{Power}}},
  \bibinfo{author}{\bibfnamefont{A.~R.} \bibnamefont{{Jenkins}}},
  \bibinfo{author}{\bibfnamefont{C.~S.} \bibnamefont{{Frenk}}},
  \bibinfo{author}{\bibfnamefont{S.~D.~M.} \bibnamefont{{White}}},
  \bibinfo{author}{\bibfnamefont{V.}~\bibnamefont{{Springel}}},
  \bibinfo{author}{\bibfnamefont{J.}~\bibnamefont{{Stadel}}}, \bibnamefont{and}
  \bibinfo{author}{\bibfnamefont{T.~R.} \bibnamefont{{Quinn}}},
  \bibinfo{journal}{Mon. Not. R. Astron. Soc.}
  \textbf{\bibinfo{volume}{349}}, \bibinfo{pages}{1039} (\bibinfo{year}{2004}).

\bibitem[{\citenamefont{{Bullock} et~al.}(2001)\citenamefont{{Bullock},
  {Kolatt}, {Sigad}, {Somerville}, {Kravtsov}, {Klypin}, {Primack}, and
  {Dekel}}}]{bullock2001}
\bibinfo{author}{\bibfnamefont{J.~S.} \bibnamefont{{Bullock}}},
  \bibinfo{author}{\bibfnamefont{T.~S.} \bibnamefont{{Kolatt}}},
  \bibinfo{author}{\bibfnamefont{Y.}~\bibnamefont{{Sigad}}},
  \bibinfo{author}{\bibfnamefont{R.~S.} \bibnamefont{{Somerville}}},
  \bibinfo{author}{\bibfnamefont{A.~V.} \bibnamefont{{Kravtsov}}},
  \bibinfo{author}{\bibfnamefont{A.~A.} \bibnamefont{{Klypin}}},
  \bibinfo{author}{\bibfnamefont{J.~R.} \bibnamefont{{Primack}}},
  \bibnamefont{and} \bibinfo{author}{\bibfnamefont{A.}~\bibnamefont{{Dekel}}},
  \bibinfo{journal}{Mon. Not. R. Astron. Soc.}
  \textbf{\bibinfo{volume}{321}}, \bibinfo{pages}{559} (\bibinfo{year}{2001}).

\bibitem[{\citenamefont{{Lu} et~al.}(2006)\citenamefont{{Lu}, {Mo}, {Katz}, and
  {Weinberg}}}]{lu2006}
\bibinfo{author}{\bibfnamefont{Y.}~\bibnamefont{{Lu}}},
  \bibinfo{author}{\bibfnamefont{H.~J.} \bibnamefont{{Mo}}},
  \bibinfo{author}{\bibfnamefont{N.}~\bibnamefont{{Katz}}}, \bibnamefont{and}
  \bibinfo{author}{\bibfnamefont{M.~D.} \bibnamefont{{Weinberg}}},
  \bibinfo{journal}{Mon. Not. R. Astron. Soc.}
  \textbf{\bibinfo{volume}{368}}, \bibinfo{pages}{1931} (\bibinfo{year}{2006}).

\bibitem[{\citenamefont{{Neto} et~al.}(2007)\citenamefont{{Neto}, {Gao},
  {Bett}, {Cole}, {Navarro}, {Frenk}, {White}, {Springel}, and
  {Jenkins}}}]{neto2007}
\bibinfo{author}{\bibfnamefont{A.~F.} \bibnamefont{{Neto}}},
  \bibinfo{author}{\bibfnamefont{L.}~\bibnamefont{{Gao}}},
  \bibinfo{author}{\bibfnamefont{P.}~\bibnamefont{{Bett}}},
  \bibinfo{author}{\bibfnamefont{S.}~\bibnamefont{{Cole}}},
  \bibinfo{author}{\bibfnamefont{J.~F.} \bibnamefont{{Navarro}}},
  \bibinfo{author}{\bibfnamefont{C.~S.} \bibnamefont{{Frenk}}},
  \bibinfo{author}{\bibfnamefont{S.~D.~M.} \bibnamefont{{White}}},
  \bibinfo{author}{\bibfnamefont{V.}~\bibnamefont{{Springel}}},
  \bibnamefont{and}
  \bibinfo{author}{\bibfnamefont{A.}~\bibnamefont{{Jenkins}}},
  \bibinfo{journal}{Mon. Not. R. Astron. Soc.}
  \textbf{\bibinfo{volume}{381}}, \bibinfo{pages}{1450} (\bibinfo{year}{2007}).

\bibitem[{\citenamefont{{Strigari}
  et~al.}(2008{\natexlab{a}})\citenamefont{{Strigari}, {Bullock}, {Kaplinghat},
  {Simon}, {Geha}, {Willman}, and {Walker}}}]{strigari_nat2008}
\bibinfo{author}{\bibfnamefont{L.~E.} \bibnamefont{{Strigari}}},
  \bibinfo{author}{\bibfnamefont{J.~S.} \bibnamefont{{Bullock}}},
  \bibinfo{author}{\bibfnamefont{M.}~\bibnamefont{{Kaplinghat}}},
  \bibinfo{author}{\bibfnamefont{J.~D.} \bibnamefont{{Simon}}},
  \bibinfo{author}{\bibfnamefont{M.}~\bibnamefont{{Geha}}},
  \bibinfo{author}{\bibfnamefont{B.}~\bibnamefont{{Willman}}},
  \bibnamefont{and} \bibinfo{author}{\bibfnamefont{M.~G.}
  \bibnamefont{{Walker}}}, \bibinfo{journal}{\nat}
  \textbf{\bibinfo{volume}{454}}, \bibinfo{pages}{1096}
  (\bibinfo{year}{2008}{\natexlab{a}}).

\bibitem[{\citenamefont{{Macci{\`o}} et~al.}(2007)\citenamefont{{Macci{\`o}},
  {Dutton}, {van den Bosch}, {Moore}, {Potter}, and {Stadel}}}]{maccio2007}
\bibinfo{author}{\bibfnamefont{A.~V.} \bibnamefont{{Macci{\`o}}}},
  \bibinfo{author}{\bibfnamefont{A.~A.} \bibnamefont{{Dutton}}},
  \bibinfo{author}{\bibfnamefont{F.~C.} \bibnamefont{{van den Bosch}}},
  \bibinfo{author}{\bibfnamefont{B.}~\bibnamefont{{Moore}}},
  \bibinfo{author}{\bibfnamefont{D.}~\bibnamefont{{Potter}}}, \bibnamefont{and}
  \bibinfo{author}{\bibfnamefont{J.}~\bibnamefont{{Stadel}}},
  \bibinfo{journal}{Mon. Not. R. Astron. Soc.}
  \textbf{\bibinfo{volume}{378}}, \bibinfo{pages}{55} (\bibinfo{year}{2007}).

\bibitem[{\citenamefont{{Macci{\`o}} et~al.}(2008)\citenamefont{{Macci{\`o}},
  {Dutton}, and {van den Bosch}}}]{maccio2008}
\bibinfo{author}{\bibfnamefont{A.~V.} \bibnamefont{{Macci{\`o}}}},
  \bibinfo{author}{\bibfnamefont{A.~A.} \bibnamefont{{Dutton}}},
  \bibnamefont{and} \bibinfo{author}{\bibfnamefont{F.~C.} \bibnamefont{{van den
  Bosch}}}, \bibinfo{journal}{Mon. Not. R. Astron. Soc.}
  \textbf{\bibinfo{volume}{391}}, \bibinfo{pages}{1940} (\bibinfo{year}{2008}).

\bibitem[{\citenamefont{{Flores} et~al.}(1986)\citenamefont{{Flores},
  {Blumenthal}, {Dekel}, and {Primack}}}]{flores1986}
\bibinfo{author}{\bibfnamefont{R.~A.} \bibnamefont{{Flores}}},
  \bibinfo{author}{\bibfnamefont{G.~R.} \bibnamefont{{Blumenthal}}},
  \bibinfo{author}{\bibfnamefont{A.}~\bibnamefont{{Dekel}}}, \bibnamefont{and}
  \bibinfo{author}{\bibfnamefont{J.~R.} \bibnamefont{{Primack}}},
  \bibinfo{journal}{\nat} \textbf{\bibinfo{volume}{323}}, \bibinfo{pages}{781}
  (\bibinfo{year}{1986}).

\bibitem[{\citenamefont{{Cen}}(2001)}]{cen2001}
\bibinfo{author}{\bibfnamefont{R.}~\bibnamefont{{Cen}}},
  \bibinfo{journal}{\apj} \textbf{\bibinfo{volume}{546}}, \bibinfo{pages}{L77}
  (\bibinfo{year}{2001}).

\bibitem[{\citenamefont{{Binney} and {Tremaine}}(2008)}]{binney2008}
\bibinfo{author}{\bibfnamefont{J.}~\bibnamefont{{Binney}}} \bibnamefont{and}
  \bibinfo{author}{\bibfnamefont{S.}~\bibnamefont{{Tremaine}}},
  \emph{\bibinfo{title}{{Galactic Dynamics}}} (\bibinfo{publisher}{Princeton University Press, Princeton}, \bibinfo{year}{2008}).

\bibitem[{\citenamefont{{Diemand} and {Moore}}(2009)}]{diemand2009}
\bibinfo{author}{\bibfnamefont{J.}~\bibnamefont{{Diemand}}} \bibnamefont{and}
  \bibinfo{author}{\bibfnamefont{B.}~\bibnamefont{{Moore}}},
  \bibinfo{journal}{arXiv:0906.4340 [Adv. Sci. Lett. (to be published)]}.

\bibitem[{\citenamefont{{Peter} et.~al.}(2010)\citenamefont{{Peter}, {Moody}, 
  and {Kamionkowski}}}]{peter2010}
\bibinfo{author}{\bibfnamefont{A.~H.~G.}~\bibnamefont{Peter}},
  \bibinfo{author}{\bibfnamefont{C.~E.}~\bibnamefont{Moody}} \bibnamefont{and}
  \bibinfo{author}{\bibfnamefont{M.}~\bibnamefont{Kamionkowski}},
  \bibinfo{journal}{arXiv:1003.0419}. 

\bibitem[{\citenamefont{{Oguri} et~al.}(2003)\citenamefont{{Oguri},
  {Takahashi}, {Ohno}, and {Kotake}}}]{oguri2003}
\bibinfo{author}{\bibfnamefont{M.}~\bibnamefont{{Oguri}}},
  \bibinfo{author}{\bibfnamefont{K.}~\bibnamefont{{Takahashi}}},
  \bibinfo{author}{\bibfnamefont{H.}~\bibnamefont{{Ohno}}}, \bibnamefont{and}
  \bibinfo{author}{\bibfnamefont{K.}~\bibnamefont{{Kotake}}},
  \bibinfo{journal}{Astrophys. J.} \textbf{\bibinfo{volume}{597}},
  \bibinfo{pages}{645} (\bibinfo{year}{2003}).

\bibitem[{\citenamefont{{Gong} and {Chen}}(2008)}]{gong2008}
\bibinfo{author}{\bibfnamefont{Y.}~\bibnamefont{{Gong}}} \bibnamefont{and}
  \bibinfo{author}{\bibfnamefont{X.}~\bibnamefont{{Chen}}},
  \bibinfo{journal}{\prd} \textbf{\bibinfo{volume}{77}},
  \bibinfo{pages}{103511} (\bibinfo{year}{2008}).

\bibitem[{\citenamefont{{Dodelson}}(2003)}]{dodelson2003}
\bibinfo{author}{\bibfnamefont{S.}~\bibnamefont{{Dodelson}}},
  \emph{\bibinfo{title}{{Modern Cosmology}}} (\bibinfo{publisher}{Academic Press, Amsterdam}, \bibinfo{year}{2003}).

\bibitem[{\citenamefont{{Bahcall} et~al.}(2003)\citenamefont{{Bahcall}, {Dong},
  {Bode}, {Kim}, {Annis}, {McKay}, {Hansen}, {Schroeder}, {Gunn}, {Ostriker}
  et~al.}}]{bahcall2003a}
\bibinfo{author}{\bibfnamefont{N.~A.} \bibnamefont{{Bahcall}}},
  \bibinfo{author}{\bibfnamefont{F.}~\bibnamefont{{Dong}}},
  \bibinfo{author}{\bibfnamefont{P.}~\bibnamefont{{Bode}}},
  \bibinfo{author}{\bibfnamefont{R.}~\bibnamefont{{Kim}}},
  \bibinfo{author}{\bibfnamefont{J.}~\bibnamefont{{Annis}}},
  \bibinfo{author}{\bibfnamefont{T.~A.} \bibnamefont{{McKay}}},
  \bibinfo{author}{\bibfnamefont{S.}~\bibnamefont{{Hansen}}},
  \bibinfo{author}{\bibfnamefont{J.}~\bibnamefont{{Schroeder}}},
  \bibinfo{author}{\bibfnamefont{J.}~\bibnamefont{{Gunn}}},
  \bibinfo{author}{\bibfnamefont{J.~P.} \bibnamefont{{Ostriker}}}
  \bibnamefont{et~al.}, \bibinfo{journal}{\apj} \textbf{\bibinfo{volume}{585}},
  \bibinfo{pages}{182} (\bibinfo{year}{2003}).

\bibitem[{\citenamefont{{Rines} et~al.}(2007)\citenamefont{{Rines}, {Diaferio},
  and {Natarajan}}}]{rines2007}
\bibinfo{author}{\bibfnamefont{K.}~\bibnamefont{{Rines}}},
  \bibinfo{author}{\bibfnamefont{A.}~\bibnamefont{{Diaferio}}},
  \bibnamefont{and}
  \bibinfo{author}{\bibfnamefont{P.}~\bibnamefont{{Natarajan}}},
  \bibinfo{journal}{\apj} \textbf{\bibinfo{volume}{657}}, \bibinfo{pages}{183}
  (\bibinfo{year}{2007}).

\bibitem[{\citenamefont{{Rines} et~al.}(2008)\citenamefont{{Rines}, {Diaferio},
  and {Natarajan}}}]{rines2008b}
\bibinfo{author}{\bibfnamefont{K.}~\bibnamefont{{Rines}}},
  \bibinfo{author}{\bibfnamefont{A.}~\bibnamefont{{Diaferio}}},
  \bibnamefont{and}
  \bibinfo{author}{\bibfnamefont{P.}~\bibnamefont{{Natarajan}}},
  \bibinfo{journal}{\apj} \textbf{\bibinfo{volume}{679}}, \bibinfo{pages}{L1}
  (\bibinfo{year}{2008}).

\bibitem[{\citenamefont{{Rozo} et~al.}(2010)\citenamefont{{Rozo}, {Wechsler},
  {Rykoff}, {Annis}, {Becker}, {Evrard}, {Frieman}, {Hansen}, {Hao}, {Johnston}
  et~al.}}]{rozo2010}
\bibinfo{author}{\bibfnamefont{E.}~\bibnamefont{{Rozo}}},
  \bibinfo{author}{\bibfnamefont{R.~H.} \bibnamefont{{Wechsler}}},
  \bibinfo{author}{\bibfnamefont{E.~S.} \bibnamefont{{Rykoff}}},
  \bibinfo{author}{\bibfnamefont{J.~T.} \bibnamefont{{Annis}}},
  \bibinfo{author}{\bibfnamefont{M.~R.} \bibnamefont{{Becker}}},
  \bibinfo{author}{\bibfnamefont{A.~E.} \bibnamefont{{Evrard}}},
  \bibinfo{author}{\bibfnamefont{J.~A.} \bibnamefont{{Frieman}}},
  \bibinfo{author}{\bibfnamefont{S.~M.} \bibnamefont{{Hansen}}},
  \bibinfo{author}{\bibfnamefont{J.}~\bibnamefont{{Hao}}},
  \bibinfo{author}{\bibfnamefont{D.~E.} \bibnamefont{{Johnston}}}
  \bibnamefont{et~al.}, \bibinfo{journal}{\apj} \textbf{\bibinfo{volume}{708}},
  \bibinfo{pages}{645} (\bibinfo{year}{2010}).

\bibitem[{\citenamefont{{Vikhlinin}
  et~al.}(2009{\natexlab{b}})\citenamefont{{Vikhlinin}, {Burenin}, {Ebeling},
  {Forman}, {Hornstrup}, {Jones}, {Kravtsov}, {Murray}, {Nagai}, {Quintana}
  et~al.}}]{vikhlinin2009a}
\bibinfo{author}{\bibfnamefont{A.}~\bibnamefont{{Vikhlinin}}},
  \bibinfo{author}{\bibfnamefont{R.~A.} \bibnamefont{{Burenin}}},
  \bibinfo{author}{\bibfnamefont{H.}~\bibnamefont{{Ebeling}}},
  \bibinfo{author}{\bibfnamefont{W.~R.} \bibnamefont{{Forman}}},
  \bibinfo{author}{\bibfnamefont{A.}~\bibnamefont{{Hornstrup}}},
  \bibinfo{author}{\bibfnamefont{C.}~\bibnamefont{{Jones}}},
  \bibinfo{author}{\bibfnamefont{A.~V.} \bibnamefont{{Kravtsov}}},
  \bibinfo{author}{\bibfnamefont{S.~S.} \bibnamefont{{Murray}}},
  \bibinfo{author}{\bibfnamefont{D.}~\bibnamefont{{Nagai}}},
  \bibinfo{author}{\bibfnamefont{H.}~\bibnamefont{{Quintana}}}
  \bibnamefont{et~al.}, \bibinfo{journal}{\apj} \textbf{\bibinfo{volume}{692}},
  \bibinfo{pages}{1033} (\bibinfo{year}{2009}{\natexlab{b}}).

\bibitem[{\citenamefont{{Carlstrom} et~al.}(2002)\citenamefont{{Carlstrom},
  {Holder}, and {Reese}}}]{carlstrom2002}
\bibinfo{author}{\bibfnamefont{J.~E.} \bibnamefont{{Carlstrom}}},
  \bibinfo{author}{\bibfnamefont{G.~P.} \bibnamefont{{Holder}}},
  \bibnamefont{and} \bibinfo{author}{\bibfnamefont{E.~D.}
  \bibnamefont{{Reese}}}, \bibinfo{journal}{Annu. Rev. Astron. Astrophys.}
  \textbf{\bibinfo{volume}{40}}, \bibinfo{pages}{643} (\bibinfo{year}{2002}).

\bibitem[{\citenamefont{{Dunkley} et~al.}(2009)\citenamefont{{Dunkley},
  {Komatsu}, {Nolta}, {Spergel}, {Larson}, {Hinshaw}, {Page}, {Bennett},
  {Gold}, {Jarosik} et~al.}}]{dunkley2009}
\bibinfo{author}{\bibfnamefont{J.}~\bibnamefont{{Dunkley}}},
  \bibinfo{author}{\bibfnamefont{E.}~\bibnamefont{{Komatsu}}},
  \bibinfo{author}{\bibfnamefont{M.~R.} \bibnamefont{{Nolta}}},
  \bibinfo{author}{\bibfnamefont{D.~N.} \bibnamefont{{Spergel}}},
  \bibinfo{author}{\bibfnamefont{D.}~\bibnamefont{{Larson}}},
  \bibinfo{author}{\bibfnamefont{G.}~\bibnamefont{{Hinshaw}}},
  \bibinfo{author}{\bibfnamefont{L.}~\bibnamefont{{Page}}},
  \bibinfo{author}{\bibfnamefont{C.~L.} \bibnamefont{{Bennett}}},
  \bibinfo{author}{\bibfnamefont{B.}~\bibnamefont{{Gold}}},
  \bibinfo{author}{\bibfnamefont{N.}~\bibnamefont{{Jarosik}}}
  \bibnamefont{et~al.}, \bibinfo{journal}{Astrophys. J. Suppl. Ser.}
  \textbf{\bibinfo{volume}{180}}, \bibinfo{pages}{306} (\bibinfo{year}{2009}).

\bibitem[{\citenamefont{{Heymans} et~al.}(2005)\citenamefont{{Heymans},
  {Brown}, {Barden}, {Caldwell}, {Jahnke}, {Peng}, {Rix}, {Taylor}, {Beckwith},
  {Bell} et~al.}}]{heymans2005}
\bibinfo{author}{\bibfnamefont{C.}~\bibnamefont{{Heymans}}},
  \bibinfo{author}{\bibfnamefont{M.~L.} \bibnamefont{{Brown}}},
  \bibinfo{author}{\bibfnamefont{M.}~\bibnamefont{{Barden}}},
  \bibinfo{author}{\bibfnamefont{J.~A.~R.} \bibnamefont{{Caldwell}}},
  \bibinfo{author}{\bibfnamefont{K.}~\bibnamefont{{Jahnke}}},
  \bibinfo{author}{\bibfnamefont{C.~Y.} \bibnamefont{{Peng}}},
  \bibinfo{author}{\bibfnamefont{H.}~\bibnamefont{{Rix}}},
  \bibinfo{author}{\bibfnamefont{A.}~\bibnamefont{{Taylor}}},
  \bibinfo{author}{\bibfnamefont{S.~V.~W.} \bibnamefont{{Beckwith}}},
  \bibinfo{author}{\bibfnamefont{E.~F.} \bibnamefont{{Bell}}}
  \bibnamefont{et~al.}, \bibinfo{journal}{Mon. Not. R. Astron. Soc.}
  \textbf{\bibinfo{volume}{361}}, \bibinfo{pages}{160} (\bibinfo{year}{2005}).

\bibitem[{\citenamefont{{Seljak} et~al.}(2005)\citenamefont{{Seljak},
  {Makarov}, {Mandelbaum}, {Hirata}, {Padmanabhan}, {McDonald}, {Blanton},
  {Tegmark}, {Bahcall}, and {Brinkmann}}}]{seljak2005}
\bibinfo{author}{\bibfnamefont{U.}~\bibnamefont{{Seljak}}},
  \bibinfo{author}{\bibfnamefont{A.}~\bibnamefont{{Makarov}}},
  \bibinfo{author}{\bibfnamefont{R.}~\bibnamefont{{Mandelbaum}}},
  \bibinfo{author}{\bibfnamefont{C.~M.} \bibnamefont{{Hirata}}},
  \bibinfo{author}{\bibfnamefont{N.}~\bibnamefont{{Padmanabhan}}},
  \bibinfo{author}{\bibfnamefont{P.}~\bibnamefont{{McDonald}}},
  \bibinfo{author}{\bibfnamefont{M.~R.} \bibnamefont{{Blanton}}},
  \bibinfo{author}{\bibfnamefont{M.}~\bibnamefont{{Tegmark}}},
  \bibinfo{author}{\bibfnamefont{N.~A.} \bibnamefont{{Bahcall}}},
  \bibnamefont{and}
  \bibinfo{author}{\bibfnamefont{J.}~\bibnamefont{{Brinkmann}}},
  \bibinfo{journal}{\prd} \textbf{\bibinfo{volume}{71}},
  \bibinfo{pages}{043511} (\bibinfo{year}{2005}).

\bibitem[{\citenamefont{{Tegmark} et~al.}(2006)\citenamefont{{Tegmark},
  {Eisenstein}, {Strauss}, {Weinberg}, {Blanton}, {Frieman}, {Fukugita},
  {Gunn}, {Hamilton}, {Knapp} et~al.}}]{tegmark2006}
\bibinfo{author}{\bibfnamefont{M.}~\bibnamefont{{Tegmark}}},
  \bibinfo{author}{\bibfnamefont{D.~J.} \bibnamefont{{Eisenstein}}},
  \bibinfo{author}{\bibfnamefont{M.~A.} \bibnamefont{{Strauss}}},
  \bibinfo{author}{\bibfnamefont{D.~H.} \bibnamefont{{Weinberg}}},
  \bibinfo{author}{\bibfnamefont{M.~R.} \bibnamefont{{Blanton}}},
  \bibinfo{author}{\bibfnamefont{J.~A.} \bibnamefont{{Frieman}}},
  \bibinfo{author}{\bibfnamefont{M.}~\bibnamefont{{Fukugita}}},
  \bibinfo{author}{\bibfnamefont{J.~E.} \bibnamefont{{Gunn}}},
  \bibinfo{author}{\bibfnamefont{A.~J.~S.} \bibnamefont{{Hamilton}}},
  \bibinfo{author}{\bibfnamefont{G.~R.} \bibnamefont{{Knapp}}}
  \bibnamefont{et~al.}, \bibinfo{journal}{\prd} \textbf{\bibinfo{volume}{74}},
  \bibinfo{pages}{123507} (\bibinfo{year}{2006}).

\bibitem[{\citenamefont{{Mandelbaum} and {Seljak}}(2007)}]{mandelbaum2007}
\bibinfo{author}{\bibfnamefont{R.}~\bibnamefont{{Mandelbaum}}}
  \bibnamefont{and} \bibinfo{author}{\bibfnamefont{U.}~\bibnamefont{{Seljak}}},
  \bibinfo{journal}{J. Cosmol. Astropart. Phys.}
  \textbf{\bibinfo{volume}{06}}, \bibinfo{pages}{024} (\bibinfo{year}{2007}).

\bibitem[{\citenamefont{{Massey} et~al.}(2007)\citenamefont{{Massey}, {Rhodes},
  {Leauthaud}, {Capak}, {Ellis}, {Koekemoer}, {R{\'e}fr{\'e}gier}, {Scoville},
  {Taylor}, {Albert} et~al.}}]{massey2007}
\bibinfo{author}{\bibfnamefont{R.}~\bibnamefont{{Massey}}},
  \bibinfo{author}{\bibfnamefont{J.}~\bibnamefont{{Rhodes}}},
  \bibinfo{author}{\bibfnamefont{A.}~\bibnamefont{{Leauthaud}}},
  \bibinfo{author}{\bibfnamefont{P.}~\bibnamefont{{Capak}}},
  \bibinfo{author}{\bibfnamefont{R.}~\bibnamefont{{Ellis}}},
  \bibinfo{author}{\bibfnamefont{A.}~\bibnamefont{{Koekemoer}}},
  \bibinfo{author}{\bibfnamefont{A.}~\bibnamefont{{R{\'e}fr{\'e}gier}}},
  \bibinfo{author}{\bibfnamefont{N.}~\bibnamefont{{Scoville}}},
  \bibinfo{author}{\bibfnamefont{J.~E.} \bibnamefont{{Taylor}}},
  \bibinfo{author}{\bibfnamefont{J.}~\bibnamefont{{Albert}}}
  \bibnamefont{et~al.}, \bibinfo{journal}{Astrophys. J. Suppl. Ser.}
  \textbf{\bibinfo{volume}{172}}, \bibinfo{pages}{239} (\bibinfo{year}{2007}).

\bibitem[{\citenamefont{{Padmanabhan} et~al.}(2007)\citenamefont{{Padmanabhan},
  {Schlegel}, {Seljak}, {Makarov}, {Bahcall}, {Blanton}, {Brinkmann},
  {Eisenstein}, {Finkbeiner}, {Gunn} et~al.}}]{padmanabhan2007}
\bibinfo{author}{\bibfnamefont{N.}~\bibnamefont{{Padmanabhan}}},
  \bibinfo{author}{\bibfnamefont{D.~J.} \bibnamefont{{Schlegel}}},
  \bibinfo{author}{\bibfnamefont{U.}~\bibnamefont{{Seljak}}},
  \bibinfo{author}{\bibfnamefont{A.}~\bibnamefont{{Makarov}}},
  \bibinfo{author}{\bibfnamefont{N.~A.} \bibnamefont{{Bahcall}}},
  \bibinfo{author}{\bibfnamefont{M.~R.} \bibnamefont{{Blanton}}},
  \bibinfo{author}{\bibfnamefont{J.}~\bibnamefont{{Brinkmann}}},
  \bibinfo{author}{\bibfnamefont{D.~J.} \bibnamefont{{Eisenstein}}},
  \bibinfo{author}{\bibfnamefont{D.~P.} \bibnamefont{{Finkbeiner}}},
  \bibinfo{author}{\bibfnamefont{J.~E.} \bibnamefont{{Gunn}}}
  \bibnamefont{et~al.}, \bibinfo{journal}{Mon. Not. R. Astron. Soc.}
  \textbf{\bibinfo{volume}{378}}, \bibinfo{pages}{852} (\bibinfo{year}{2007}).

\bibitem[{\citenamefont{{Percival} et~al.}(2007)}]{percival2007}
\bibinfo{author}{\bibfnamefont{W.~J.} \bibnamefont{{Percival}}}
  \bibnamefont{et~al.}, \bibinfo{journal}{Astrophys. J.}
  \textbf{\bibinfo{volume}{657}}, \bibinfo{pages}{645} (\bibinfo{year}{2007}).

\bibitem[{\citenamefont{{Press} and {Schechter}}(1974)}]{press1974}
\bibinfo{author}{\bibfnamefont{W.~H.} \bibnamefont{{Press}}} \bibnamefont{and}
  \bibinfo{author}{\bibfnamefont{P.}~\bibnamefont{{Schechter}}},
  \bibinfo{journal}{\apj} \textbf{\bibinfo{volume}{187}}, \bibinfo{pages}{425}
  (\bibinfo{year}{1974}).

\bibitem[{\citenamefont{{Sheth} and {Tormen}}(2002)}]{sheth2002}
\bibinfo{author}{\bibfnamefont{R.~K.} \bibnamefont{{Sheth}}} \bibnamefont{and}
  \bibinfo{author}{\bibfnamefont{G.}~\bibnamefont{{Tormen}}},
  \bibinfo{journal}{Mon. Not. R. Astron. Soc.}
  \textbf{\bibinfo{volume}{329}}, \bibinfo{pages}{61} (\bibinfo{year}{2002}).

\bibitem[{\citenamefont{{Tinker} et~al.}(2008)\citenamefont{{Tinker},
  {Kravtsov}, {Klypin}, {Abazajian}, {Warren}, {Yepes}, {Gottl{\"o}ber}, and
  {Holz}}}]{tinker2008}
\bibinfo{author}{\bibfnamefont{J.}~\bibnamefont{{Tinker}}},
  \bibinfo{author}{\bibfnamefont{A.~V.} \bibnamefont{{Kravtsov}}},
  \bibinfo{author}{\bibfnamefont{A.}~\bibnamefont{{Klypin}}},
  \bibinfo{author}{\bibfnamefont{K.}~\bibnamefont{{Abazajian}}},
  \bibinfo{author}{\bibfnamefont{M.}~\bibnamefont{{Warren}}},
  \bibinfo{author}{\bibfnamefont{G.}~\bibnamefont{{Yepes}}},
  \bibinfo{author}{\bibfnamefont{S.}~\bibnamefont{{Gottl{\"o}ber}}},
  \bibnamefont{and} \bibinfo{author}{\bibfnamefont{D.~E.}
  \bibnamefont{{Holz}}}, \bibinfo{journal}{\apj}
  \textbf{\bibinfo{volume}{688}}, \bibinfo{pages}{709} (\bibinfo{year}{2008}).

\bibitem[{\citenamefont{{Cunha} et~al.}(2009)\citenamefont{{Cunha}, {Huterer},
  and {Frieman}}}]{cunha2009}
\bibinfo{author}{\bibfnamefont{C.}~\bibnamefont{{Cunha}}},
  \bibinfo{author}{\bibfnamefont{D.}~\bibnamefont{{Huterer}}},
  \bibnamefont{and} \bibinfo{author}{\bibfnamefont{J.~A.}
  \bibnamefont{{Frieman}}}, \bibinfo{journal}{\prd}
  \textbf{\bibinfo{volume}{80}}, \bibinfo{pages}{063532}
  (\bibinfo{year}{2009}), \eprint{0904.1589}.

\bibitem[{\citenamefont{{Zehavi} et~al.}(2005)}]{zehavi2005}
\bibinfo{author}{\bibfnamefont{I.}~\bibnamefont{{Zehavi}}}
  \bibnamefont{et~al.}, \bibinfo{journal}{Astrophys. J.}
  \textbf{\bibinfo{volume}{630}}, \bibinfo{pages}{1} (\bibinfo{year}{2005}).

\bibitem[{\citenamefont{{Bond} and {Szalay}}(1983)}]{bond1983}
\bibinfo{author}{\bibfnamefont{J.~R.} \bibnamefont{{Bond}}} \bibnamefont{and}
  \bibinfo{author}{\bibfnamefont{A.~S.} \bibnamefont{{Szalay}}},
  \bibinfo{journal}{\apj} \textbf{\bibinfo{volume}{274}}, \bibinfo{pages}{443}
  (\bibinfo{year}{1983}).

\bibitem[{\citenamefont{{Seljak}}(2000)}]{seljak2000}
\bibinfo{author}{\bibfnamefont{U.}~\bibnamefont{{Seljak}}},
  \bibinfo{journal}{Mon. Not. R. Astron. Soc.}
  \textbf{\bibinfo{volume}{318}}, \bibinfo{pages}{203} (\bibinfo{year}{2000}).

\bibitem[{\citenamefont{{Berlind} and {Weinberg}}(2002)}]{berlind2002}
\bibinfo{author}{\bibfnamefont{A.~A.} \bibnamefont{{Berlind}}}
  \bibnamefont{and} \bibinfo{author}{\bibfnamefont{D.~H.}
  \bibnamefont{{Weinberg}}}, \bibinfo{journal}{\apj}
  \textbf{\bibinfo{volume}{575}}, \bibinfo{pages}{587} (\bibinfo{year}{2002}).

\bibitem[{\citenamefont{{Zheng} et~al.}(2002)\citenamefont{{Zheng}, {Tinker},
  {Weinberg}, and {Berlind}}}]{zheng2002}
\bibinfo{author}{\bibfnamefont{Z.}~\bibnamefont{{Zheng}}},
  \bibinfo{author}{\bibfnamefont{J.~L.} \bibnamefont{{Tinker}}},
  \bibinfo{author}{\bibfnamefont{D.~H.} \bibnamefont{{Weinberg}}},
  \bibnamefont{and} \bibinfo{author}{\bibfnamefont{A.~A.}
  \bibnamefont{{Berlind}}}, \bibinfo{journal}{\apj}
  \textbf{\bibinfo{volume}{575}}, \bibinfo{pages}{617} (\bibinfo{year}{2002}).

\bibitem[{\citenamefont{{Zheng} et~al.}(2005)\citenamefont{{Zheng}, {Berlind},
  {Weinberg}, {Benson}, {Baugh}, {Cole}, {Dav{\'e}}, {Frenk}, {Katz}, and
  {Lacey}}}]{zheng2005}
\bibinfo{author}{\bibfnamefont{Z.}~\bibnamefont{{Zheng}}},
  \bibinfo{author}{\bibfnamefont{A.~A.} \bibnamefont{{Berlind}}},
  \bibinfo{author}{\bibfnamefont{D.~H.} \bibnamefont{{Weinberg}}},
  \bibinfo{author}{\bibfnamefont{A.~J.} \bibnamefont{{Benson}}},
  \bibinfo{author}{\bibfnamefont{C.~M.} \bibnamefont{{Baugh}}},
  \bibinfo{author}{\bibfnamefont{S.}~\bibnamefont{{Cole}}},
  \bibinfo{author}{\bibfnamefont{R.}~\bibnamefont{{Dav{\'e}}}},
  \bibinfo{author}{\bibfnamefont{C.~S.} \bibnamefont{{Frenk}}},
  \bibinfo{author}{\bibfnamefont{N.}~\bibnamefont{{Katz}}}, \bibnamefont{and}
  \bibinfo{author}{\bibfnamefont{C.~G.} \bibnamefont{{Lacey}}},
  \bibinfo{journal}{\apj} \textbf{\bibinfo{volume}{633}}, \bibinfo{pages}{791}
  (\bibinfo{year}{2005}).

\bibitem[{\citenamefont{{Zheng} and {Weinberg}}(2007)}]{zheng2007}
\bibinfo{author}{\bibfnamefont{Z.}~\bibnamefont{{Zheng}}} \bibnamefont{and}
  \bibinfo{author}{\bibfnamefont{D.~H.} \bibnamefont{{Weinberg}}},
  \bibinfo{journal}{\apj} \textbf{\bibinfo{volume}{659}}, \bibinfo{pages}{1}
  (\bibinfo{year}{2007}).

\bibitem[{\citenamefont{{Abazajian} et~al.}(2005)\citenamefont{{Abazajian},
  {Zheng}, {Zehavi}, {Weinberg}, {Frieman}, {Berlind}, {Blanton}, {Bahcall},
  {Brinkmann}, {Schneider} et~al.}}]{abazajian2005a}
\bibinfo{author}{\bibfnamefont{K.}~\bibnamefont{{Abazajian}}},
  \bibinfo{author}{\bibfnamefont{Z.}~\bibnamefont{{Zheng}}},
  \bibinfo{author}{\bibfnamefont{I.}~\bibnamefont{{Zehavi}}},
  \bibinfo{author}{\bibfnamefont{D.~H.} \bibnamefont{{Weinberg}}},
  \bibinfo{author}{\bibfnamefont{J.~A.} \bibnamefont{{Frieman}}},
  \bibinfo{author}{\bibfnamefont{A.~A.} \bibnamefont{{Berlind}}},
  \bibinfo{author}{\bibfnamefont{M.~R.} \bibnamefont{{Blanton}}},
  \bibinfo{author}{\bibfnamefont{N.~A.} \bibnamefont{{Bahcall}}},
  \bibinfo{author}{\bibfnamefont{J.}~\bibnamefont{{Brinkmann}}},
  \bibinfo{author}{\bibfnamefont{D.~P.} \bibnamefont{{Schneider}}}
  \bibnamefont{et~al.}, \bibinfo{journal}{\apj} \textbf{\bibinfo{volume}{625}},
  \bibinfo{pages}{613} (\bibinfo{year}{2005}).

\bibitem[{\citenamefont{{White} and {Padmanabhan}}(2009)}]{white2009}
\bibinfo{author}{\bibfnamefont{M.}~\bibnamefont{{White}}} \bibnamefont{and}
  \bibinfo{author}{\bibfnamefont{N.}~\bibnamefont{{Padmanabhan}}},
  \bibinfo{journal}{Mon. Not. R. Astron. Soc.}
  \textbf{\bibinfo{volume}{395}}, \bibinfo{pages}{2381} (\bibinfo{year}{2009}).

\bibitem[{\citenamefont{{Mandelbaum}
  et~al.}(2006{\natexlab{a}})\citenamefont{{Mandelbaum}, {Seljak}, {Kauffmann},
  {Hirata}, and {Brinkmann}}}]{mandelbaum2006}
\bibinfo{author}{\bibfnamefont{R.}~\bibnamefont{{Mandelbaum}}},
  \bibinfo{author}{\bibfnamefont{U.}~\bibnamefont{{Seljak}}},
  \bibinfo{author}{\bibfnamefont{G.}~\bibnamefont{{Kauffmann}}},
  \bibinfo{author}{\bibfnamefont{C.~M.} \bibnamefont{{Hirata}}},
  \bibnamefont{and}
  \bibinfo{author}{\bibfnamefont{J.}~\bibnamefont{{Brinkmann}}},
  \bibinfo{journal}{Mon. Not. R. Astron. Soc.}
  \textbf{\bibinfo{volume}{368}}, \bibinfo{pages}{715}
  (\bibinfo{year}{2006}{\natexlab{a}}).

\bibitem[{\citenamefont{{Mandelbaum}
  et~al.}(2006{\natexlab{b}})\citenamefont{{Mandelbaum}, {Seljak}, {Cool},
  {Blanton}, {Hirata}, and {Brinkmann}}}]{mandelbaum2006a}
\bibinfo{author}{\bibfnamefont{R.}~\bibnamefont{{Mandelbaum}}},
  \bibinfo{author}{\bibfnamefont{U.}~\bibnamefont{{Seljak}}},
  \bibinfo{author}{\bibfnamefont{R.~J.} \bibnamefont{{Cool}}},
  \bibinfo{author}{\bibfnamefont{M.}~\bibnamefont{{Blanton}}},
  \bibinfo{author}{\bibfnamefont{C.~M.} \bibnamefont{{Hirata}}},
  \bibnamefont{and}
  \bibinfo{author}{\bibfnamefont{J.}~\bibnamefont{{Brinkmann}}},
  \bibinfo{journal}{Mon. Not. R. Astron. Soc.}
  \textbf{\bibinfo{volume}{372}}, \bibinfo{pages}{758}
  (\bibinfo{year}{2006}{\natexlab{b}}).

\bibitem[{\citenamefont{{LSST Science Collaborations: Paul A.~Abell}
  et~al.}(2009)\citenamefont{{LSST Science Collaborations: Paul A.~Abell},
  {Allison}, {Anderson}, {Andrew}, {Angel}, {Armus}, {Arnett}, {Asztalos},
  {Axelrod}, {Bailey} et~al.}}]{lsst2009}
\bibinfo{author}{\bibnamefont{{LSST Science Collaborations: Paul A.~Abell}}},
  \bibinfo{author}{\bibfnamefont{J.}~\bibnamefont{{Allison}}},
  \bibinfo{author}{\bibfnamefont{S.~F.} \bibnamefont{{Anderson}}},
  \bibinfo{author}{\bibfnamefont{J.~R.} \bibnamefont{{Andrew}}},
  \bibinfo{author}{\bibfnamefont{J.~R.~P.} \bibnamefont{{Angel}}},
  \bibinfo{author}{\bibfnamefont{L.}~\bibnamefont{{Armus}}},
  \bibinfo{author}{\bibfnamefont{D.}~\bibnamefont{{Arnett}}},
  \bibinfo{author}{\bibfnamefont{S.~J.} \bibnamefont{{Asztalos}}},
  \bibinfo{author}{\bibfnamefont{T.~S.} \bibnamefont{{Axelrod}}},
  \bibinfo{author}{\bibfnamefont{S.}~\bibnamefont{{Bailey}}}
  \bibnamefont{et~al.}, \bibinfo{journal}{arXiv:0912.0201}.

\bibitem[{\citenamefont{{Vegetti} et~al.}(2009)\citenamefont{{Vegetti},
  {Koopmans}, {Bolton}, {Treu}, and {Gavazzi}}}]{vegetti2009c}
\bibinfo{author}{\bibfnamefont{S.}~\bibnamefont{{Vegetti}}},
  \bibinfo{author}{\bibfnamefont{L.~V.~E.} \bibnamefont{{Koopmans}}},
  \bibinfo{author}{\bibfnamefont{A.}~\bibnamefont{{Bolton}}},
  \bibinfo{author}{\bibfnamefont{T.}~\bibnamefont{{Treu}}}, \bibnamefont{and}
  \bibinfo{author}{\bibfnamefont{R.}~\bibnamefont{{Gavazzi}}},
  \bibinfo{journal}{arXiv:0910.0760}.

\bibitem[{\citenamefont{{Duffy} et~al.}(2008)\citenamefont{{Duffy}, {Schaye},
  {Kay}, and {Dalla Vecchia}}}]{duffy2008}
\bibinfo{author}{\bibfnamefont{A.~R.} \bibnamefont{{Duffy}}},
  \bibinfo{author}{\bibfnamefont{J.}~\bibnamefont{{Schaye}}},
  \bibinfo{author}{\bibfnamefont{S.~T.} \bibnamefont{{Kay}}}, \bibnamefont{and}
  \bibinfo{author}{\bibfnamefont{C.}~\bibnamefont{{Dalla Vecchia}}},
  \bibinfo{journal}{Mon. Not. R. Astron. Soc.}
  \textbf{\bibinfo{volume}{390}}, \bibinfo{pages}{L64} (\bibinfo{year}{2008}).

\bibitem[{\citenamefont{{Gao} et~al.}(2008)\citenamefont{{Gao}, {Navarro},
  {Cole}, {Frenk}, {White}, {Springel}, {Jenkins}, and {Neto}}}]{gao2008}
\bibinfo{author}{\bibfnamefont{L.}~\bibnamefont{{Gao}}},
  \bibinfo{author}{\bibfnamefont{J.~F.} \bibnamefont{{Navarro}}},
  \bibinfo{author}{\bibfnamefont{S.}~\bibnamefont{{Cole}}},
  \bibinfo{author}{\bibfnamefont{C.~S.} \bibnamefont{{Frenk}}},
  \bibinfo{author}{\bibfnamefont{S.~D.~M.} \bibnamefont{{White}}},
  \bibinfo{author}{\bibfnamefont{V.}~\bibnamefont{{Springel}}},
  \bibinfo{author}{\bibfnamefont{A.}~\bibnamefont{{Jenkins}}},
  \bibnamefont{and} \bibinfo{author}{\bibfnamefont{A.~F.}
  \bibnamefont{{Neto}}}, \bibinfo{journal}{Mon. Not. R. Astron. Soc.}
  \textbf{\bibinfo{volume}{387}}, \bibinfo{pages}{536} (\bibinfo{year}{2008}).

\bibitem[{\citenamefont{{Rudd} et~al.}(2008)\citenamefont{{Rudd}, {Zentner},
  and {Kravtsov}}}]{rudd2008}
\bibinfo{author}{\bibfnamefont{D.~H.} \bibnamefont{{Rudd}}},
  \bibinfo{author}{\bibfnamefont{A.~R.} \bibnamefont{{Zentner}}},
  \bibnamefont{and} \bibinfo{author}{\bibfnamefont{A.~V.}
  \bibnamefont{{Kravtsov}}}, \bibinfo{journal}{\apj}
  \textbf{\bibinfo{volume}{672}}, \bibinfo{pages}{19} (\bibinfo{year}{2008}).

\bibitem[{\citenamefont{{Rines} and {Diaferio}}(2006)}]{rines2006}
\bibinfo{author}{\bibfnamefont{K.}~\bibnamefont{{Rines}}} \bibnamefont{and}
  \bibinfo{author}{\bibfnamefont{A.}~\bibnamefont{{Diaferio}}},
  \bibinfo{journal}{Astron. J.} \textbf{\bibinfo{volume}{132}},
  \bibinfo{pages}{1275} (\bibinfo{year}{2006}).

\bibitem[{\citenamefont{{Buote} et~al.}(2007)\citenamefont{{Buote},
  {Gastaldello}, {Humphrey}, {Zappacosta}, {Bullock}, {Brighenti}, and
  {Mathews}}}]{buote2007}
\bibinfo{author}{\bibfnamefont{D.~A.} \bibnamefont{{Buote}}},
  \bibinfo{author}{\bibfnamefont{F.}~\bibnamefont{{Gastaldello}}},
  \bibinfo{author}{\bibfnamefont{P.~J.} \bibnamefont{{Humphrey}}},
  \bibinfo{author}{\bibfnamefont{L.}~\bibnamefont{{Zappacosta}}},
  \bibinfo{author}{\bibfnamefont{J.~S.} \bibnamefont{{Bullock}}},
  \bibinfo{author}{\bibfnamefont{F.}~\bibnamefont{{Brighenti}}},
  \bibnamefont{and} \bibinfo{author}{\bibfnamefont{W.~G.}
  \bibnamefont{{Mathews}}}, \bibinfo{journal}{\apj}
  \textbf{\bibinfo{volume}{664}}, \bibinfo{pages}{123} (\bibinfo{year}{2007}).

\bibitem[{\citenamefont{{Comerford} and {Natarajan}}(2007)}]{comerford2007}
\bibinfo{author}{\bibfnamefont{J.~M.} \bibnamefont{{Comerford}}}
  \bibnamefont{and}
  \bibinfo{author}{\bibfnamefont{P.}~\bibnamefont{{Natarajan}}},
  \bibinfo{journal}{Mon. Not. R. Astron. Soc.}
  \textbf{\bibinfo{volume}{379}}, \bibinfo{pages}{190} (\bibinfo{year}{2007}).

\bibitem[{\citenamefont{{Mandelbaum} et~al.}(2008)\citenamefont{{Mandelbaum},
  {Seljak}, and {Hirata}}}]{mandelbaum2008}
\bibinfo{author}{\bibfnamefont{R.}~\bibnamefont{{Mandelbaum}}},
  \bibinfo{author}{\bibfnamefont{U.}~\bibnamefont{{Seljak}}}, \bibnamefont{and}
  \bibinfo{author}{\bibfnamefont{C.~M.} \bibnamefont{{Hirata}}},
  \bibinfo{journal}{J. Cosmol. Astropart. Phys.}
  \textbf{\bibinfo{volume}{08}}, \bibinfo{pages}{006} (\bibinfo{year}{2008}).

\bibitem[{\citenamefont{{Spergel} et~al.}(2003)\citenamefont{{Spergel},
  {Verde}, {Peiris}, {Komatsu}, {Nolta}, {Bennett}, {Halpern}, {Hinshaw},
  {Jarosik}, {Kogut} et~al.}}]{spergel2003}
\bibinfo{author}{\bibfnamefont{D.~N.} \bibnamefont{{Spergel}}},
  \bibinfo{author}{\bibfnamefont{L.}~\bibnamefont{{Verde}}},
  \bibinfo{author}{\bibfnamefont{H.~V.} \bibnamefont{{Peiris}}},
  \bibinfo{author}{\bibfnamefont{E.}~\bibnamefont{{Komatsu}}},
  \bibinfo{author}{\bibfnamefont{M.~R.} \bibnamefont{{Nolta}}},
  \bibinfo{author}{\bibfnamefont{C.~L.} \bibnamefont{{Bennett}}},
  \bibinfo{author}{\bibfnamefont{M.}~\bibnamefont{{Halpern}}},
  \bibinfo{author}{\bibfnamefont{G.}~\bibnamefont{{Hinshaw}}},
  \bibinfo{author}{\bibfnamefont{N.}~\bibnamefont{{Jarosik}}},
  \bibinfo{author}{\bibfnamefont{A.}~\bibnamefont{{Kogut}}}
  \bibnamefont{et~al.}, \bibinfo{journal}{Astrophys. J. Suppl. Ser.}
  \textbf{\bibinfo{volume}{148}}, \bibinfo{pages}{175} (\bibinfo{year}{2003}).

\bibitem[{\citenamefont{{Spergel} et~al.}(2007)}]{spergel2007}
\bibinfo{author}{\bibfnamefont{D.~N.} \bibnamefont{{Spergel}}}
  \bibnamefont{et~al.}, \bibinfo{journal}{Astrophys. J. Suppl. Ser.}
  \textbf{\bibinfo{volume}{170}}, \bibinfo{pages}{377} (\bibinfo{year}{2007}).

\bibitem[{\citenamefont{{Hennawi} et~al.}(2007)\citenamefont{{Hennawi},
  {Dalal}, {Bode}, and {Ostriker}}}]{hennawi2007}
\bibinfo{author}{\bibfnamefont{J.~F.} \bibnamefont{{Hennawi}}},
  \bibinfo{author}{\bibfnamefont{N.}~\bibnamefont{{Dalal}}},
  \bibinfo{author}{\bibfnamefont{P.}~\bibnamefont{{Bode}}}, \bibnamefont{and}
  \bibinfo{author}{\bibfnamefont{J.~P.} \bibnamefont{{Ostriker}}},
  \bibinfo{journal}{\apj} \textbf{\bibinfo{volume}{654}}, \bibinfo{pages}{714}
  (\bibinfo{year}{2007}).

\bibitem[{\citenamefont{{van de Ven} et~al.}(2009)\citenamefont{{van de Ven},
  {Mandelbaum}, and {Keeton}}}]{vandeven2009}
\bibinfo{author}{\bibfnamefont{G.}~\bibnamefont{{van de Ven}}},
  \bibinfo{author}{\bibfnamefont{R.}~\bibnamefont{{Mandelbaum}}},
  \bibnamefont{and} \bibinfo{author}{\bibfnamefont{C.~R.}
  \bibnamefont{{Keeton}}}, \bibinfo{journal}{Mon. Not. R. Astron. Soc.}
  \textbf{\bibinfo{volume}{398}}, \bibinfo{pages}{607} (\bibinfo{year}{2009}).

\bibitem[{\citenamefont{{Mandelbaum} et~al.}(2009)\citenamefont{{Mandelbaum},
  {van de Ven}, and {Keeton}}}]{mandelbaum2009}
\bibinfo{author}{\bibfnamefont{R.}~\bibnamefont{{Mandelbaum}}},
  \bibinfo{author}{\bibfnamefont{G.}~\bibnamefont{{van de Ven}}},
  \bibnamefont{and} \bibinfo{author}{\bibfnamefont{C.~R.}
  \bibnamefont{{Keeton}}}, \bibinfo{journal}{Mon. Not. R. Astron. Soc.}
  \textbf{\bibinfo{volume}{398}}, \bibinfo{pages}{635} (\bibinfo{year}{2009}).

\bibitem[{\citenamefont{{Ichiki} et~al.}(2004)\citenamefont{{Ichiki}, {Oguri},
  and {Takahashi}}}]{ichiki2004}
\bibinfo{author}{\bibfnamefont{K.}~\bibnamefont{{Ichiki}}},
  \bibinfo{author}{\bibfnamefont{M.}~\bibnamefont{{Oguri}}}, \bibnamefont{and}
  \bibinfo{author}{\bibfnamefont{K.}~\bibnamefont{{Takahashi}}},
  \bibinfo{journal}{Phys. Rev. Lett.} \textbf{\bibinfo{volume}{93}},
  \bibinfo{pages}{071302} (\bibinfo{year}{2004}).

\bibitem[{\citenamefont{{Zentner} and {Walker}}(2002)}]{zentner2002}
\bibinfo{author}{\bibfnamefont{A.~R.} \bibnamefont{{Zentner}}}
  \bibnamefont{and} \bibinfo{author}{\bibfnamefont{T.~P.}
  \bibnamefont{{Walker}}}, \bibinfo{journal}{\prd}
  \textbf{\bibinfo{volume}{65}}, \bibinfo{pages}{063506}
  (\bibinfo{year}{2002}).

\bibitem[{\citenamefont{{Dalal} and {Kochanek}}(2002)}]{dalal2002}
\bibinfo{author}{\bibfnamefont{N.}~\bibnamefont{{Dalal}}} \bibnamefont{and}
  \bibinfo{author}{\bibfnamefont{C.~S.} \bibnamefont{{Kochanek}}},
  \bibinfo{journal}{\apj} \textbf{\bibinfo{volume}{572}}, \bibinfo{pages}{25}
  (\bibinfo{year}{2002}).

\bibitem[{\citenamefont{{Keeton} and {Moustakas}}(2009)}]{keeton2009}
\bibinfo{author}{\bibfnamefont{C.~R.} \bibnamefont{{Keeton}}} \bibnamefont{and}
  \bibinfo{author}{\bibfnamefont{L.~A.} \bibnamefont{{Moustakas}}},
  \bibinfo{journal}{\apj} \textbf{\bibinfo{volume}{699}}, \bibinfo{pages}{1720}
  (\bibinfo{year}{2009}).

\bibitem[{\citenamefont{{Koopmans} et~al.}(2009)\citenamefont{{Koopmans},
  {Barnabe}, {Bolton}, {Bradac}, {Ciotti}, {Congdon}, {Czoske}, {Dye},
  {Dutton}, {Elliasdottir} et~al.}}]{koopmans2009}
\bibinfo{author}{\bibfnamefont{L.~V.~E.} \bibnamefont{{Koopmans}}},
  \bibinfo{author}{\bibfnamefont{M.}~\bibnamefont{{Barnabe}}},
  \bibinfo{author}{\bibfnamefont{A.}~\bibnamefont{{Bolton}}},
  \bibinfo{author}{\bibfnamefont{M.}~\bibnamefont{{Bradac}}},
  \bibinfo{author}{\bibfnamefont{L.}~\bibnamefont{{Ciotti}}},
  \bibinfo{author}{\bibfnamefont{A.}~\bibnamefont{{Congdon}}},
  \bibinfo{author}{\bibfnamefont{O.}~\bibnamefont{{Czoske}}},
  \bibinfo{author}{\bibfnamefont{S.}~\bibnamefont{{Dye}}},
  \bibinfo{author}{\bibfnamefont{A.}~\bibnamefont{{Dutton}}},
  \bibinfo{author}{\bibfnamefont{A.}~\bibnamefont{{Elliasdottir}}}
  \bibnamefont{et~al.}, \bibinfo{journal}{arXiv:0902.3186}.


\bibitem[{\citenamefont{{Marshall} et~al.}(2009)\citenamefont{{Marshall},
  {Auger}, {Bartlett}, {Bradac}, {Cooray}, {Dalal}, {Dobler}, {Fassnacht},
  {Jain}, {Keeton} et~al.}}]{marshall2009}
\bibinfo{author}{\bibfnamefont{P.~J.} \bibnamefont{{Marshall}}},
  \bibinfo{author}{\bibfnamefont{M.}~\bibnamefont{{Auger}}},
  \bibinfo{author}{\bibfnamefont{J.~G.} \bibnamefont{{Bartlett}}},
  \bibinfo{author}{\bibfnamefont{M.}~\bibnamefont{{Bradac}}},
  \bibinfo{author}{\bibfnamefont{A.}~\bibnamefont{{Cooray}}},
  \bibinfo{author}{\bibfnamefont{N.}~\bibnamefont{{Dalal}}},
  \bibinfo{author}{\bibfnamefont{G.}~\bibnamefont{{Dobler}}},
  \bibinfo{author}{\bibfnamefont{C.~D.} \bibnamefont{{Fassnacht}}},
  \bibinfo{author}{\bibfnamefont{B.}~\bibnamefont{{Jain}}},
  \bibinfo{author}{\bibfnamefont{C.~R.} \bibnamefont{{Keeton}}}
  \bibnamefont{et~al.}, \bibinfo{journal}{arXiv:0902.2963}.

\bibitem[{\citenamefont{{Moustakas} et~al.}(2009)\citenamefont{{Moustakas},
  {Abazajian}, {Benson}, {Bolton}, {Bullock}, {Chen}, {Cheng}, {Coe},
  {Congdon}, {Dalal} et~al.}}]{moustakas2009}
\bibinfo{author}{\bibfnamefont{L.~A.} \bibnamefont{{Moustakas}}},
  \bibinfo{author}{\bibfnamefont{K.}~\bibnamefont{{Abazajian}}},
  \bibinfo{author}{\bibfnamefont{A.}~\bibnamefont{{Benson}}},
  \bibinfo{author}{\bibfnamefont{A.~S.} \bibnamefont{{Bolton}}},
  \bibinfo{author}{\bibfnamefont{J.~S.} \bibnamefont{{Bullock}}},
  \bibinfo{author}{\bibfnamefont{J.}~\bibnamefont{{Chen}}},
  \bibinfo{author}{\bibfnamefont{E.}~\bibnamefont{{Cheng}}},
  \bibinfo{author}{\bibfnamefont{D.}~\bibnamefont{{Coe}}},
  \bibinfo{author}{\bibfnamefont{A.~B.} \bibnamefont{{Congdon}}},
  \bibinfo{author}{\bibfnamefont{N.}~\bibnamefont{{Dalal}}}
  \bibnamefont{et~al.}, \bibinfo{journal}{arXiv:0902.3219}.

\bibitem[{\citenamefont{{Gilmore} et~al.}(2007)\citenamefont{{Gilmore},
  {Wilkinson}, {Wyse}, {Kleyna}, {Koch}, {Evans}, and {Grebel}}}]{gilmore2007}
\bibinfo{author}{\bibfnamefont{G.}~\bibnamefont{{Gilmore}}},
  \bibinfo{author}{\bibfnamefont{M.~I.} \bibnamefont{{Wilkinson}}},
  \bibinfo{author}{\bibfnamefont{R.~F.~G.} \bibnamefont{{Wyse}}},
  \bibinfo{author}{\bibfnamefont{J.~T.} \bibnamefont{{Kleyna}}},
  \bibinfo{author}{\bibfnamefont{A.}~\bibnamefont{{Koch}}},
  \bibinfo{author}{\bibfnamefont{N.~W.} \bibnamefont{{Evans}}},
  \bibnamefont{and} \bibinfo{author}{\bibfnamefont{E.~K.}
  \bibnamefont{{Grebel}}}, \bibinfo{journal}{\apj}
  \textbf{\bibinfo{volume}{663}}, \bibinfo{pages}{948} (\bibinfo{year}{2007}).

\bibitem[{\citenamefont{{Strigari} et~al.}(2007)\citenamefont{{Strigari},
  {Bullock}, and {Kaplinghat}}}]{strigari2007d}
\bibinfo{author}{\bibfnamefont{L.~E.} \bibnamefont{{Strigari}}},
  \bibinfo{author}{\bibfnamefont{J.~S.} \bibnamefont{{Bullock}}},
  \bibnamefont{and}
  \bibinfo{author}{\bibfnamefont{M.}~\bibnamefont{{Kaplinghat}}},
  \bibinfo{journal}{\apj} \textbf{\bibinfo{volume}{657}}, \bibinfo{pages}{L1}
  (\bibinfo{year}{2007}).

\bibitem[{\citenamefont{{Strigari}
  et~al.}(2008{\natexlab{b}})\citenamefont{{Strigari}, {Koushiappas},
  {Bullock}, {Kaplinghat}, {Simon}, {Geha}, and {Willman}}}]{strigari2008}
\bibinfo{author}{\bibfnamefont{L.~E.} \bibnamefont{{Strigari}}},
  \bibinfo{author}{\bibfnamefont{S.~M.} \bibnamefont{{Koushiappas}}},
  \bibinfo{author}{\bibfnamefont{J.~S.} \bibnamefont{{Bullock}}},
  \bibinfo{author}{\bibfnamefont{M.}~\bibnamefont{{Kaplinghat}}},
  \bibinfo{author}{\bibfnamefont{J.~D.} \bibnamefont{{Simon}}},
  \bibinfo{author}{\bibfnamefont{M.}~\bibnamefont{{Geha}}}, \bibnamefont{and}
  \bibinfo{author}{\bibfnamefont{B.}~\bibnamefont{{Willman}}},
  \bibinfo{journal}{\apj} \textbf{\bibinfo{volume}{678}}, \bibinfo{pages}{614}
  (\bibinfo{year}{2008}{\natexlab{b}}).

\bibitem[{\citenamefont{{Walker} et~al.}(2009)\citenamefont{{Walker}, {Mateo},
  {Olszewski}, {Pe{\~n}arrubia}, {Wyn Evans}, and {Gilmore}}}]{walker2009}
\bibinfo{author}{\bibfnamefont{M.~G.} \bibnamefont{{Walker}}},
  \bibinfo{author}{\bibfnamefont{M.}~\bibnamefont{{Mateo}}},
  \bibinfo{author}{\bibfnamefont{E.~W.} \bibnamefont{{Olszewski}}},
  \bibinfo{author}{\bibfnamefont{J.}~\bibnamefont{{Pe{\~n}arrubia}}},
  \bibinfo{author}{\bibfnamefont{N.}~\bibnamefont{{Wyn Evans}}},
  \bibnamefont{and}
  \bibinfo{author}{\bibfnamefont{G.}~\bibnamefont{{Gilmore}}},
  \bibinfo{journal}{\apj} \textbf{\bibinfo{volume}{704}}, \bibinfo{pages}{1274}
  (\bibinfo{year}{2009}).

\bibitem[{\citenamefont{{Kuzio de Naray} et~al.}(2006)\citenamefont{{Kuzio de
  Naray}, {McGaugh}, {de Blok}, and {Bosma}}}]{kuzio2006}
\bibinfo{author}{\bibfnamefont{R.}~\bibnamefont{{Kuzio de Naray}}},
  \bibinfo{author}{\bibfnamefont{S.~S.} \bibnamefont{{McGaugh}}},
  \bibinfo{author}{\bibfnamefont{W.~J.~G.} \bibnamefont{{de Blok}}},
  \bibnamefont{and} \bibinfo{author}{\bibfnamefont{A.}~\bibnamefont{{Bosma}}},
  \bibinfo{journal}{Astrophys. J. Suppl. Ser.} \textbf{\bibinfo{volume}{165}},
  \bibinfo{pages}{461} (\bibinfo{year}{2006}).

\bibitem[{\citenamefont{{Bailin} et~al.}(2007)\citenamefont{{Bailin}, {Simon},
  {Bolatto}, {Gibson}, and {Power}}}]{bailin2007}
\bibinfo{author}{\bibfnamefont{J.}~\bibnamefont{{Bailin}}},
  \bibinfo{author}{\bibfnamefont{J.~D.} \bibnamefont{{Simon}}},
  \bibinfo{author}{\bibfnamefont{A.~D.} \bibnamefont{{Bolatto}}},
  \bibinfo{author}{\bibfnamefont{B.~K.} \bibnamefont{{Gibson}}},
  \bibnamefont{and} \bibinfo{author}{\bibfnamefont{C.}~\bibnamefont{{Power}}},
  \bibinfo{journal}{\apj} \textbf{\bibinfo{volume}{667}}, \bibinfo{pages}{191}
  (\bibinfo{year}{2007}).

\bibitem[{\citenamefont{{de Blok} et~al.}(2008)\citenamefont{{de Blok},
  {Walter}, {Brinks}, {Trachternach}, {Oh}, and {Kennicutt}}}]{deblok2008}
\bibinfo{author}{\bibfnamefont{W.~J.~G.} \bibnamefont{{de Blok}}},
  \bibinfo{author}{\bibfnamefont{F.}~\bibnamefont{{Walter}}},
  \bibinfo{author}{\bibfnamefont{E.}~\bibnamefont{{Brinks}}},
  \bibinfo{author}{\bibfnamefont{C.}~\bibnamefont{{Trachternach}}},
  \bibinfo{author}{\bibfnamefont{S.-H.} \bibnamefont{{Oh}}}, \bibnamefont{and}
  \bibinfo{author}{\bibfnamefont{R.~C.} \bibnamefont{{Kennicutt}}},
  \bibinfo{journal}{Astron. J.} \textbf{\bibinfo{volume}{136}},
  \bibinfo{pages}{2648} (\bibinfo{year}{2008}).

\bibitem[{\citenamefont{{Oh} et~al.}(2008)\citenamefont{{Oh}, {de Blok},
  {Walter}, {Brinks}, and {Kennicutt}}}]{oh2008}
\bibinfo{author}{\bibfnamefont{S.}~\bibnamefont{{Oh}}},
  \bibinfo{author}{\bibfnamefont{W.~J.~G.} \bibnamefont{{de Blok}}},
  \bibinfo{author}{\bibfnamefont{F.}~\bibnamefont{{Walter}}},
  \bibinfo{author}{\bibfnamefont{E.}~\bibnamefont{{Brinks}}}, \bibnamefont{and}
  \bibinfo{author}{\bibfnamefont{R.~C.} \bibnamefont{{Kennicutt}}},
  \bibinfo{journal}{Astron. J.} \textbf{\bibinfo{volume}{136}},
  \bibinfo{pages}{2761} (\bibinfo{year}{2008}).

\bibitem[{\citenamefont{{Trott} et~al.}(2010)\citenamefont{{Trott}, {Treu},
  {Koopmans}, and {Webster}}}]{trott2009}
\bibinfo{author}{\bibfnamefont{C.~M.} \bibnamefont{{Trott}}},
  \bibinfo{author}{\bibfnamefont{T.}~\bibnamefont{{Treu}}},
  \bibinfo{author}{\bibfnamefont{L.~V.~E.} \bibnamefont{{Koopmans}}},
  \bibnamefont{and} \bibinfo{author}{\bibfnamefont{R.~L.}
  \bibnamefont{{Webster}}}, \bibinfo{journal}{Mon. Not. R. Astron. Soc.} 
  \textbf{\bibinfo{volume}{401}} \bibinfo{pages}{1540} (\bibinfo{year}{2010}).

\bibitem[{\citenamefont{{Treu} and {Koopmans}}(2004)}]{treu2004}
\bibinfo{author}{\bibfnamefont{T.}~\bibnamefont{{Treu}}} \bibnamefont{and}
  \bibinfo{author}{\bibfnamefont{L.~V.~E.} \bibnamefont{{Koopmans}}},
  \bibinfo{journal}{Astrophys. J.} \textbf{\bibinfo{volume}{611}},
  \bibinfo{pages}{739} (\bibinfo{year}{2004}).

\bibitem[{\citenamefont{{Padmanabhan} et~al.}(2004)\citenamefont{{Padmanabhan},
  {Seljak}, {Strauss}, {Blanton}, {Kauffmann}, {Schlegel}, {Tremonti},
  {Bahcall}, {Bernardi}, {Brinkmann} et~al.}}]{padmanabhan2004}
\bibinfo{author}{\bibfnamefont{N.}~\bibnamefont{{Padmanabhan}}},
  \bibinfo{author}{\bibfnamefont{U.}~\bibnamefont{{Seljak}}},
  \bibinfo{author}{\bibfnamefont{M.~A.} \bibnamefont{{Strauss}}},
  \bibinfo{author}{\bibfnamefont{M.~R.} \bibnamefont{{Blanton}}},
  \bibinfo{author}{\bibfnamefont{G.}~\bibnamefont{{Kauffmann}}},
  \bibinfo{author}{\bibfnamefont{D.~J.} \bibnamefont{{Schlegel}}},
  \bibinfo{author}{\bibfnamefont{C.}~\bibnamefont{{Tremonti}}},
  \bibinfo{author}{\bibfnamefont{N.~A.} \bibnamefont{{Bahcall}}},
  \bibinfo{author}{\bibfnamefont{M.}~\bibnamefont{{Bernardi}}},
  \bibinfo{author}{\bibfnamefont{J.}~\bibnamefont{{Brinkmann}}}
  \bibnamefont{et~al.}, \bibinfo{journal}{New Astron.}
  \textbf{\bibinfo{volume}{9}}, \bibinfo{pages}{329} (\bibinfo{year}{2004}).

\bibitem[{\citenamefont{{Dekel} et~al.}(2005)\citenamefont{{Dekel}, {Stoehr},
  {Mamon}, {Cox}, {Novak}, and {Primack}}}]{dekel2005}
\bibinfo{author}{\bibfnamefont{A.}~\bibnamefont{{Dekel}}},
  \bibinfo{author}{\bibfnamefont{F.}~\bibnamefont{{Stoehr}}},
  \bibinfo{author}{\bibfnamefont{G.~A.} \bibnamefont{{Mamon}}},
  \bibinfo{author}{\bibfnamefont{T.~J.} \bibnamefont{{Cox}}},
  \bibinfo{author}{\bibfnamefont{G.~S.} \bibnamefont{{Novak}}},
  \bibnamefont{and} \bibinfo{author}{\bibfnamefont{J.~R.}
  \bibnamefont{{Primack}}}, \bibinfo{journal}{\nat}
  \textbf{\bibinfo{volume}{437}}, \bibinfo{pages}{707} (\bibinfo{year}{2005}).

\bibitem[{\citenamefont{{Humphrey} et~al.}(2006)\citenamefont{{Humphrey},
  {Buote}, {Gastaldello}, {Zappacosta}, {Bullock}, {Brighenti}, and
  {Mathews}}}]{humphrey2006}
\bibinfo{author}{\bibfnamefont{P.~J.} \bibnamefont{{Humphrey}}},
  \bibinfo{author}{\bibfnamefont{D.~A.} \bibnamefont{{Buote}}},
  \bibinfo{author}{\bibfnamefont{F.}~\bibnamefont{{Gastaldello}}},
  \bibinfo{author}{\bibfnamefont{L.}~\bibnamefont{{Zappacosta}}},
  \bibinfo{author}{\bibfnamefont{J.~S.} \bibnamefont{{Bullock}}},
  \bibinfo{author}{\bibfnamefont{F.}~\bibnamefont{{Brighenti}}},
  \bibnamefont{and} \bibinfo{author}{\bibfnamefont{W.~G.}
  \bibnamefont{{Mathews}}}, \bibinfo{journal}{\apj}
  \textbf{\bibinfo{volume}{646}}, \bibinfo{pages}{899} (\bibinfo{year}{2006}).

\bibitem[{\citenamefont{{Schulz} et~al.}(2009)\citenamefont{{Schulz},
  {Mandelbaum}, and {Padmanabhan}}}]{schulz2009}
\bibinfo{author}{\bibfnamefont{A.~E.} \bibnamefont{{Schulz}}},
  \bibinfo{author}{\bibfnamefont{R.}~\bibnamefont{{Mandelbaum}}},
  \bibnamefont{and}
  \bibinfo{author}{\bibfnamefont{N.}~\bibnamefont{{Padmanabhan}}},
  \bibinfo{journal}{arXiv:0911.2260}.

\bibitem[{\citenamefont{{Comerford} et~al.}(2006)\citenamefont{{Comerford},
  {Meneghetti}, {Bartelmann}, and {Schirmer}}}]{comerford2006}
\bibinfo{author}{\bibfnamefont{J.~M.} \bibnamefont{{Comerford}}},
  \bibinfo{author}{\bibfnamefont{M.}~\bibnamefont{{Meneghetti}}},
  \bibinfo{author}{\bibfnamefont{M.}~\bibnamefont{{Bartelmann}}},
  \bibnamefont{and}
  \bibinfo{author}{\bibfnamefont{M.}~\bibnamefont{{Schirmer}}},
  \bibinfo{journal}{Astrophys. J.} \textbf{\bibinfo{volume}{642}},
  \bibinfo{pages}{39} (\bibinfo{year}{2006}).

\bibitem[{\citenamefont{{Gastaldello} et~al.}(2007)\citenamefont{{Gastaldello},
  {Buote}, {Humphrey}, {Zappacosta}, {Bullock}, {Brighenti}, and
  {Mathews}}}]{gastaldello2007}
\bibinfo{author}{\bibfnamefont{F.}~\bibnamefont{{Gastaldello}}},
  \bibinfo{author}{\bibfnamefont{D.~A.} \bibnamefont{{Buote}}},
  \bibinfo{author}{\bibfnamefont{P.~J.} \bibnamefont{{Humphrey}}},
  \bibinfo{author}{\bibfnamefont{L.}~\bibnamefont{{Zappacosta}}},
  \bibinfo{author}{\bibfnamefont{J.~S.} \bibnamefont{{Bullock}}},
  \bibinfo{author}{\bibfnamefont{F.}~\bibnamefont{{Brighenti}}},
  \bibnamefont{and} \bibinfo{author}{\bibfnamefont{W.~G.}
  \bibnamefont{{Mathews}}}, \bibinfo{journal}{\apj}
  \textbf{\bibinfo{volume}{669}}, \bibinfo{pages}{158} (\bibinfo{year}{2007}).

\bibitem[{\citenamefont{{Limousin} et~al.}(2007)\citenamefont{{Limousin},
  {Richard}, {Jullo}, {Kneib}, {Fort}, {Soucail}, {El{\'{\i}}asd{\'o}ttir},
  {Natarajan}, {Ellis}, {Smail} et~al.}}]{limousin2007}
\bibinfo{author}{\bibfnamefont{M.}~\bibnamefont{{Limousin}}},
  \bibinfo{author}{\bibfnamefont{J.}~\bibnamefont{{Richard}}},
  \bibinfo{author}{\bibfnamefont{E.}~\bibnamefont{{Jullo}}},
  \bibinfo{author}{\bibfnamefont{J.}~\bibnamefont{{Kneib}}},
  \bibinfo{author}{\bibfnamefont{B.}~\bibnamefont{{Fort}}},
  \bibinfo{author}{\bibfnamefont{G.}~\bibnamefont{{Soucail}}},
  \bibinfo{author}{\bibfnamefont{{\'A}.}~\bibnamefont{{El{\'{\i}}asd{\'o}ttir}%
}}, \bibinfo{author}{\bibfnamefont{P.}~\bibnamefont{{Natarajan}}},
  \bibinfo{author}{\bibfnamefont{R.~S.} \bibnamefont{{Ellis}}},
  \bibinfo{author}{\bibfnamefont{I.}~\bibnamefont{{Smail}}}
  \bibnamefont{et~al.}, \bibinfo{journal}{\apj} \textbf{\bibinfo{volume}{668}},
  \bibinfo{pages}{643} (\bibinfo{year}{2007}).

\bibitem[{\citenamefont{{Sand} et~al.}(2008)\citenamefont{{Sand}, {Treu},
  {Ellis}, {Smith}, and {Kneib}}}]{sand2008}
\bibinfo{author}{\bibfnamefont{D.~J.} \bibnamefont{{Sand}}},
  \bibinfo{author}{\bibfnamefont{T.}~\bibnamefont{{Treu}}},
  \bibinfo{author}{\bibfnamefont{R.~S.} \bibnamefont{{Ellis}}},
  \bibinfo{author}{\bibfnamefont{G.~P.} \bibnamefont{{Smith}}},
  \bibnamefont{and} \bibinfo{author}{\bibfnamefont{J.-P.}
  \bibnamefont{{Kneib}}}, \bibinfo{journal}{\apj}
  \textbf{\bibinfo{volume}{674}}, \bibinfo{pages}{711} (\bibinfo{year}{2008}).

\bibitem[{\citenamefont{{Oguri} et~al.}(2009)\citenamefont{{Oguri}, {Hennawi},
  {Gladders}, {Dahle}, {Natarajan}, {Dalal}, {Koester}, {Sharon}, and
  {Bayliss}}}]{oguri2009}
\bibinfo{author}{\bibfnamefont{M.}~\bibnamefont{{Oguri}}},
  \bibinfo{author}{\bibfnamefont{J.~F.} \bibnamefont{{Hennawi}}},
  \bibinfo{author}{\bibfnamefont{M.~D.} \bibnamefont{{Gladders}}},
  \bibinfo{author}{\bibfnamefont{H.}~\bibnamefont{{Dahle}}},
  \bibinfo{author}{\bibfnamefont{P.}~\bibnamefont{{Natarajan}}},
  \bibinfo{author}{\bibfnamefont{N.}~\bibnamefont{{Dalal}}},
  \bibinfo{author}{\bibfnamefont{B.~P.} \bibnamefont{{Koester}}},
  \bibinfo{author}{\bibfnamefont{K.}~\bibnamefont{{Sharon}}}, \bibnamefont{and}
  \bibinfo{author}{\bibfnamefont{M.}~\bibnamefont{{Bayliss}}},
  \bibinfo{journal}{\apj} \textbf{\bibinfo{volume}{699}}, \bibinfo{pages}{1038}
  (\bibinfo{year}{2009}).

\bibitem[{\citenamefont{{Newman} et~al.}(2009)\citenamefont{{Newman}, {Treu},
  {Ellis}, {Sand}, {Richard}, {Marshall}, {Capak}, and
  {Miyazaki}}}]{newman2009}
\bibinfo{author}{\bibfnamefont{A.~B.} \bibnamefont{{Newman}}},
  \bibinfo{author}{\bibfnamefont{T.}~\bibnamefont{{Treu}}},
  \bibinfo{author}{\bibfnamefont{R.~S.} \bibnamefont{{Ellis}}},
  \bibinfo{author}{\bibfnamefont{D.~J.} \bibnamefont{{Sand}}},
  \bibinfo{author}{\bibfnamefont{J.}~\bibnamefont{{Richard}}},
  \bibinfo{author}{\bibfnamefont{P.~J.} \bibnamefont{{Marshall}}},
  \bibinfo{author}{\bibfnamefont{P.}~\bibnamefont{{Capak}}}, \bibnamefont{and}
  \bibinfo{author}{\bibfnamefont{S.}~\bibnamefont{{Miyazaki}}},
  \bibinfo{journal}{\apj} \textbf{\bibinfo{volume}{706}}, \bibinfo{pages}{1078}
  (\bibinfo{year}{2009}).

\bibitem[{\citenamefont{{El-Zant} et~al.}(2001)\citenamefont{{El-Zant},
  {Shlosman}, and {Hoffman}}}]{elzant2001}
\bibinfo{author}{\bibfnamefont{A.}~\bibnamefont{{El-Zant}}},
  \bibinfo{author}{\bibfnamefont{I.}~\bibnamefont{{Shlosman}}},
  \bibnamefont{and}
  \bibinfo{author}{\bibfnamefont{Y.}~\bibnamefont{{Hoffman}}},
  \bibinfo{journal}{\apj} \textbf{\bibinfo{volume}{560}}, \bibinfo{pages}{636}
  (\bibinfo{year}{2001}).

\bibitem[{\citenamefont{{Romano-D{\'{\i}}az}
  et~al.}(2008)\citenamefont{{Romano-D{\'{\i}}az}, {Shlosman}, {Hoffman}, and
  {Heller}}}]{romano-diaz2008}
\bibinfo{author}{\bibfnamefont{E.}~\bibnamefont{{Romano-D{\'{\i}}az}}},
  \bibinfo{author}{\bibfnamefont{I.}~\bibnamefont{{Shlosman}}},
  \bibinfo{author}{\bibfnamefont{Y.}~\bibnamefont{{Hoffman}}},
  \bibnamefont{and} \bibinfo{author}{\bibfnamefont{C.}~\bibnamefont{{Heller}}},
  \bibinfo{journal}{\apj} \textbf{\bibinfo{volume}{685}}, \bibinfo{pages}{L105}
  (\bibinfo{year}{2008}).

\bibitem[{\citenamefont{{Abadi} et~al.}(2009)\citenamefont{{Abadi}, {Navarro},
  {Fardal}, {Babul}, and {Steinmetz}}}]{abadi2009}
\bibinfo{author}{\bibfnamefont{M.~G.} \bibnamefont{{Abadi}}},
  \bibinfo{author}{\bibfnamefont{J.~F.} \bibnamefont{{Navarro}}},
  \bibinfo{author}{\bibfnamefont{M.}~\bibnamefont{{Fardal}}},
  \bibinfo{author}{\bibfnamefont{A.}~\bibnamefont{{Babul}}}, \bibnamefont{and}
  \bibinfo{author}{\bibfnamefont{M.}~\bibnamefont{{Steinmetz}}},
  \bibinfo{journal}{arXiv:0902.2477}.

\bibitem[{\citenamefont{{Pedrosa}
  et~al.}(2009{\natexlab{a}})\citenamefont{{Pedrosa}, {Tissera}, and
  {Scannapieco}}}]{pedrosa2010}
\bibinfo{author}{\bibfnamefont{S.}~\bibnamefont{{Pedrosa}}},
  \bibinfo{author}{\bibfnamefont{P.~B.} \bibnamefont{{Tissera}}},
  \bibnamefont{and}
  \bibinfo{author}{\bibfnamefont{C.}~\bibnamefont{{Scannapieco}}},
  \bibinfo{journal}{Mon. Not. R. Astron. Soc.} \textbf{\bibinfo{volume}{402}}, \bibinfo{pages}{776}
  (\bibinfo{year}{2010}{\natexlab{a}}).

\bibitem[{\citenamefont{{Pedrosa}
  et~al.}(2009{\natexlab{b}})\citenamefont{{Pedrosa}, {Tissera}, and
  {Scannapieco}}}]{pedrosa2009a}
\bibinfo{author}{\bibfnamefont{S.}~\bibnamefont{{Pedrosa}}},
  \bibinfo{author}{\bibfnamefont{P.~B.} \bibnamefont{{Tissera}}},
  \bibnamefont{and}
  \bibinfo{author}{\bibfnamefont{C.}~\bibnamefont{{Scannapieco}}},
  \bibinfo{journal}{Mon. Not. R. Astron. Soc.}
  \textbf{\bibinfo{volume}{395}}, \bibinfo{pages}{L57}
  (\bibinfo{year}{2009}{\natexlab{b}}).

\bibitem[{\citenamefont{{Romano-D{\'{\i}}az}
  et~al.}(2009)\citenamefont{{Romano-D{\'{\i}}az}, {Shlosman}, {Heller}, and
  {Hoffman}}}]{romano-diaz2009}
\bibinfo{author}{\bibfnamefont{E.}~\bibnamefont{{Romano-D{\'{\i}}az}}},
  \bibinfo{author}{\bibfnamefont{I.}~\bibnamefont{{Shlosman}}},
  \bibinfo{author}{\bibfnamefont{C.}~\bibnamefont{{Heller}}}, \bibnamefont{and}
  \bibinfo{author}{\bibfnamefont{Y.}~\bibnamefont{{Hoffman}}},
  \bibinfo{journal}{\apj} \textbf{\bibinfo{volume}{702}}, \bibinfo{pages}{1250}
  (\bibinfo{year}{2009}).

\bibitem[{\citenamefont{{Tissera} et~al.}(2009)\citenamefont{{Tissera},
  {White}, {Pedrosa}, and {Scannapieco}}}]{tissera2009}
\bibinfo{author}{\bibfnamefont{P.~B.} \bibnamefont{{Tissera}}},
  \bibinfo{author}{\bibfnamefont{S.~D.~M.} \bibnamefont{{White}}},
  \bibinfo{author}{\bibfnamefont{S.}~\bibnamefont{{Pedrosa}}},
  \bibnamefont{and}
  \bibinfo{author}{\bibfnamefont{C.}~\bibnamefont{{Scannapieco}}},
  \bibinfo{journal}{arXiv:0911.2316}.

\bibitem[{\citenamefont{{Governato} et~al.}(2009)\citenamefont{{Governato},
  {Brook}, {Mayer}, {Brooks}, {Rhee}, {Wadsley}, {Jonsson}, {Willman},
  {Stinson}, {Quinn} et~al.}}]{governato2009}
\bibinfo{author}{\bibfnamefont{F.}~\bibnamefont{{Governato}}},
  \bibinfo{author}{\bibfnamefont{C.}~\bibnamefont{{Brook}}},
  \bibinfo{author}{\bibfnamefont{L.}~\bibnamefont{{Mayer}}},
  \bibinfo{author}{\bibfnamefont{A.}~\bibnamefont{{Brooks}}},
  \bibinfo{author}{\bibfnamefont{G.}~\bibnamefont{{Rhee}}},
  \bibinfo{author}{\bibfnamefont{J.}~\bibnamefont{{Wadsley}}},
  \bibinfo{author}{\bibfnamefont{P.}~\bibnamefont{{Jonsson}}},
  \bibinfo{author}{\bibfnamefont{B.}~\bibnamefont{{Willman}}},
  \bibinfo{author}{\bibfnamefont{G.}~\bibnamefont{{Stinson}}},
  \bibinfo{author}{\bibfnamefont{T.}~\bibnamefont{{Quinn}}}
  \bibnamefont{et~al.}, \bibinfo{journal}{arXiv:0911.2237}.

\bibitem[{\citenamefont{{Gnedin} et~al.}(2004)\citenamefont{{Gnedin},
  {Kravtsov}, {Klypin}, and {Nagai}}}]{gnedin2004}
\bibinfo{author}{\bibfnamefont{O.~Y.} \bibnamefont{{Gnedin}}},
  \bibinfo{author}{\bibfnamefont{A.~V.} \bibnamefont{{Kravtsov}}},
  \bibinfo{author}{\bibfnamefont{A.~A.} \bibnamefont{{Klypin}}},
  \bibnamefont{and} \bibinfo{author}{\bibfnamefont{D.}~\bibnamefont{{Nagai}}},
  \bibinfo{journal}{Astrophys. J.} \textbf{\bibinfo{volume}{616}},
  \bibinfo{pages}{16} (\bibinfo{year}{2004}).

\bibitem[{\citenamefont{{Gustafsson} et~al.}(2006)\citenamefont{{Gustafsson},
  {Fairbairn}, and {Sommer-Larsen}}}]{gustafsson2006}
\bibinfo{author}{\bibfnamefont{M.}~\bibnamefont{{Gustafsson}}},
  \bibinfo{author}{\bibfnamefont{M.}~\bibnamefont{{Fairbairn}}},
  \bibnamefont{and}
  \bibinfo{author}{\bibfnamefont{J.}~\bibnamefont{{Sommer-Larsen}}},
  \bibinfo{journal}{\prd} \textbf{\bibinfo{volume}{74}},
  \bibinfo{pages}{123522} (\bibinfo{year}{2006}).

\bibitem[{\citenamefont{{Annis} et~al.}(2005)\citenamefont{{Annis}, {Bridle},
  {Castander}, {Evrard}, {Fosalba}, {Frieman}, {Gaztanaga}, {Jain}, {Kravtsov},
  {Lahav} et~al.}}]{annis2005}
\bibinfo{author}{\bibfnamefont{J.}~\bibnamefont{{Annis}}},
  \bibinfo{author}{\bibfnamefont{S.}~\bibnamefont{{Bridle}}},
  \bibinfo{author}{\bibfnamefont{F.~J.} \bibnamefont{{Castander}}},
  \bibinfo{author}{\bibfnamefont{A.~E.} \bibnamefont{{Evrard}}},
  \bibinfo{author}{\bibfnamefont{P.}~\bibnamefont{{Fosalba}}},
  \bibinfo{author}{\bibfnamefont{J.~A.} \bibnamefont{{Frieman}}},
  \bibinfo{author}{\bibfnamefont{E.}~\bibnamefont{{Gaztanaga}}},
  \bibinfo{author}{\bibfnamefont{B.}~\bibnamefont{{Jain}}},
  \bibinfo{author}{\bibfnamefont{A.~V.} \bibnamefont{{Kravtsov}}},
  \bibinfo{author}{\bibfnamefont{O.}~\bibnamefont{{Lahav}}}
  \bibnamefont{et~al.}, \eprint{arXiv:astro-ph/0510195}.

\bibitem[{\citenamefont{{Kaiser} et~al.}(2002)\citenamefont{{Kaiser}, {Aussel},
  {Burke}, {Boesgaard}, {Chambers}, {Chun}, {Heasley}, {Hodapp}, {Hunt},
  {Jedicke} et~al.}}]{kaiser2002}
\bibinfo{author}{\bibfnamefont{N.}~\bibnamefont{{Kaiser}}},
  \bibinfo{author}{\bibfnamefont{H.}~\bibnamefont{{Aussel}}},
  \bibinfo{author}{\bibfnamefont{B.~E.} \bibnamefont{{Burke}}},
  \bibinfo{author}{\bibfnamefont{H.}~\bibnamefont{{Boesgaard}}},
  \bibinfo{author}{\bibfnamefont{K.}~\bibnamefont{{Chambers}}},
  \bibinfo{author}{\bibfnamefont{M.~R.} \bibnamefont{{Chun}}},
  \bibinfo{author}{\bibfnamefont{J.~N.} \bibnamefont{{Heasley}}},
  \bibinfo{author}{\bibfnamefont{K.}~\bibnamefont{{Hodapp}}},
  \bibinfo{author}{\bibfnamefont{B.}~\bibnamefont{{Hunt}}},
  \bibinfo{author}{\bibfnamefont{R.}~\bibnamefont{{Jedicke}}}
  \bibnamefont{et~al.}, in \emph{\bibinfo{booktitle}{Society of Photo-Optical
  Instrumentation Engineers (SPIE) Conference Series}}, edited by
  \bibinfo{editor}{\bibnamefont{{J.~A.~Tyson \& S.~Wolff}}}
  (\bibinfo{year}{2002}), vol. \bibinfo{volume}{4836} of
  \emph{\bibinfo{series}{Presented at the Society of Photo-Optical
  Instrumentation Engineers (SPIE) Conference}}, pp. \bibinfo{pages}{154--164}.

\bibitem[{\citenamefont{{Staniszewski}
  et~al.}(2009)\citenamefont{{Staniszewski}, {Ade}, {Aird}, {Benson}, {Bleem},
  {Carlstrom}, {Chang}, {Cho}, {Crawford}, {Crites} et~al.}}]{staniszewski2009}
\bibinfo{author}{\bibfnamefont{Z.}~\bibnamefont{{Staniszewski}}},
  \bibinfo{author}{\bibfnamefont{P.~A.~R.} \bibnamefont{{Ade}}},
  \bibinfo{author}{\bibfnamefont{K.~A.} \bibnamefont{{Aird}}},
  \bibinfo{author}{\bibfnamefont{B.~A.} \bibnamefont{{Benson}}},
  \bibinfo{author}{\bibfnamefont{L.~E.} \bibnamefont{{Bleem}}},
  \bibinfo{author}{\bibfnamefont{J.~E.} \bibnamefont{{Carlstrom}}},
  \bibinfo{author}{\bibfnamefont{C.~L.} \bibnamefont{{Chang}}},
  \bibinfo{author}{\bibfnamefont{H.}~\bibnamefont{{Cho}}},
  \bibinfo{author}{\bibfnamefont{T.~M.} \bibnamefont{{Crawford}}},
  \bibinfo{author}{\bibfnamefont{A.~T.} \bibnamefont{{Crites}}}
  \bibnamefont{et~al.}, \bibinfo{journal}{\apj} \textbf{\bibinfo{volume}{701}},
  \bibinfo{pages}{32} (\bibinfo{year}{2009}).

\bibitem[{\citenamefont{{Hincks} et~al.}(2009)\citenamefont{{Hincks},
  {Acquaviva}, {Ade}, {Aguirre}, {Amiri}, {Appel}, {Barrientos}, {Battistelli},
  {Bond}, {Brown} et~al.}}]{hincks2009}
\bibinfo{author}{\bibfnamefont{A.~D.} \bibnamefont{{Hincks}}},
  \bibinfo{author}{\bibfnamefont{V.}~\bibnamefont{{Acquaviva}}},
  \bibinfo{author}{\bibfnamefont{P.}~\bibnamefont{{Ade}}},
  \bibinfo{author}{\bibfnamefont{P.}~\bibnamefont{{Aguirre}}},
  \bibinfo{author}{\bibfnamefont{M.}~\bibnamefont{{Amiri}}},
  \bibinfo{author}{\bibfnamefont{J.~W.} \bibnamefont{{Appel}}},
  \bibinfo{author}{\bibfnamefont{L.~F.} \bibnamefont{{Barrientos}}},
  \bibinfo{author}{\bibfnamefont{E.~S.} \bibnamefont{{Battistelli}}},
  \bibinfo{author}{\bibfnamefont{J.~R.} \bibnamefont{{Bond}}},
  \bibinfo{author}{\bibfnamefont{B.}~\bibnamefont{{Brown}}}
  \bibnamefont{et~al.}, \bibinfo{journal}{arXiv:0907.0461}.

\bibitem[{\citenamefont{{Predehl} et~al.}(2006)\citenamefont{{Predehl},
  {Hasinger}, {B{\"o}hringer}, {Briel}, {Brunner}, {Churazov}, {Freyberg},
  {Friedrich}, {Kendziorra}, {Lutz} et~al.}}]{predehl2006}
\bibinfo{author}{\bibfnamefont{P.}~\bibnamefont{{Predehl}}},
  \bibinfo{author}{\bibfnamefont{G.}~\bibnamefont{{Hasinger}}},
  \bibinfo{author}{\bibfnamefont{H.}~\bibnamefont{{B{\"o}hringer}}},
  \bibinfo{author}{\bibfnamefont{U.}~\bibnamefont{{Briel}}},
  \bibinfo{author}{\bibfnamefont{H.}~\bibnamefont{{Brunner}}},
  \bibinfo{author}{\bibfnamefont{E.}~\bibnamefont{{Churazov}}},
  \bibinfo{author}{\bibfnamefont{M.}~\bibnamefont{{Freyberg}}},
  \bibinfo{author}{\bibfnamefont{P.}~\bibnamefont{{Friedrich}}},
  \bibinfo{author}{\bibfnamefont{E.}~\bibnamefont{{Kendziorra}}},
  \bibinfo{author}{\bibfnamefont{D.}~\bibnamefont{{Lutz}}}
  \bibnamefont{et~al.}, in \emph{\bibinfo{booktitle}{Society of Photo-Optical
  Instrumentation Engineers (SPIE) Conference Series}} (\bibinfo{year}{2006}),
  vol. \bibinfo{volume}{6266} of \emph{\bibinfo{series}{Presented at the
  Society of Photo-Optical Instrumentation Engineers (SPIE) Conference}}.

\end{thebibliography}

\end{document}